%% file: main-with-appendices.tex
\input{\jobname.cfg}
\documentclass[acmsmall,screen]{acmart}
\usepackage[T1]{fontenc}
\usepackage[utf8]{inputenc}
\usepackage[scaled=0.8]{beramono}

\usepackage{boolean}
\usepackage{reversion}

\newcommand{\Anonymous}{\FALSE}

\newcommand{\nonanon}[2]{\Anonymous{#1 removed for anonymity.}{#2}}

\usepackage{mybiblio}

\usepackage{enumitem}

\usepackage{mathtools}

\let\oldexists\exists
\let\exists\relax\DeclareMathOperator{\exists}{\oldexists}
\let\oldforall\forall
\let\forall\relax\DeclareMathOperator{\forall}{\oldforall}

\DeclareMathOperator{\lambdaAbs}{\lambda}
\DeclareMathOperator{\muAbs}{\mu}
\DeclareMathOperator{\nuAbs}{\nu}

\usepackage{amsthm}

\usepackage{bm}

\usepackage{mathpartir}
\let\TirName\textsc

\newcommand{\RefTirName}[1]{\hyperlink{#1}{\TirName{#1}}\xspace}

\newenvironment{mathline}
   {\abovedisplayskip 0.2em
    \belowdisplayskip 0.2em
    \begin{mathpar}}
   {\end{mathpar}}

\usepackage[capitalize,noabbrev,nameinlink]{cleveref}

\usepackage{mycomments}
\UNXXX

\usepackage{macros}

\usepackage{listings/ocaml}
\usepackage{listings/datalang}

\setcopyright{rightsretained}
\acmDOI{10.1145/3704915}
\acmJournal{PACMPL}
\acmYear{2025}
\acmVolume{9}
\acmNumber{POPL}
\acmArticle{79}
\acmMonth{1}
\received{2024-07-11}
\received[accepted]{2024-11-07}

\makeatletter
\def\@font@info#1{}
\makeatother

\begin{document}


\title{Tail Modulo Cons, \OCaml, and Relational Separation Logic}

\author{Clément Allain}
\orcid{0009-0005-2972-5181}
\affiliation{%
  \institution{Inria}
  \city{Paris}
  \country{France}
}
\email{clement.allain@inria.fr}

\author{Frédéric Bour}
\orcid{0009-0007-5268-5784}
\affiliation{%
  \institution{Tarides}
  \city{Paris}
  \country{France}
}
\email{frederic.bour@lakaban.net}

\author{Basile Clément}
\orcid{0000-0002-9126-0937}
\affiliation{%
  \institution{OCamlPro}
  \city{Paris}
  \country{France}
}
\email{bc@ocamlpro.com}

\author{François Pottier}
\orcid{0000-0002-4069-1235}
\affiliation{%
  \institution{Inria}
  \city{Paris}
  \country{France}
}
\email{francois.pottier@inria.fr}

\author{Gabriel Scherer}
\orcid{0000-0003-1758-3938}
\affiliation{%
  \institution{Inria}
  \city{Paris}
  \country{France}
}
\affiliation{%
  \institution{IRIF, Université Paris Cité}
  \city{Paris}
  \country{France}
}
\email{gabriel.scherer@inria.fr}


\begin{abstract}
    \input{abstract}
\end{abstract}

\titlenote{
\Appendices{
\textbf{Appendices included}: This is a long version of the present paper, with appendices.
}{
\textbf{Appendices missing}: A long version of this paper, with appendices, is available online.
}
}

\maketitle




\newcommand{\separate}{\FALSE}

\input{introduction}

\separate{\clearpage}{}
\input{formalization}

\separate{\clearpage}{}
\input{ocaml}

\separate{\clearpage}{}
\input{specification}

\separate{\clearpage}{}
\input{program_logic}

\separate{\clearpage}{}
\input{protocols}

\separate{\clearpage}{}
\input{proof}

\separate{\clearpage}{}
\input{simulation}

\separate{\clearpage}{}
\input{related}


%
%

\input{acknowledgments}


\bibliography{tmc}


\begin{version}{\Appendices}
\clearpage
\appendix

\begin{version}{\FALSE}
\input{republication}
\end{version}

\input{more-ocaml}

\input{ocaml4-appendix}
\end{version}
\end{document}

%% file: abstract.tex
Common functional languages incentivize tail-recursive functions, as opposed to general recursive functions that consume stack space and may not scale to large inputs.
This distinction occasionally requires writing functions in a tail-recursive style that may be more complex and slower than the natural, non-tail-recursive definition.

This work describes our implementation of the \emph{tail modulo constructor} (TMC) transformation in the \OCaml compiler, an optimization that provides stack-efficiency for a larger class of functions --- tail-recursive \emph{modulo constructors} --- which includes in particular the natural definition of \ocaml{List.map} and many similar recursive data-constructing functions.

We prove the correctness of this program transformation in a simplified setting --- a small untyped calculus --- that captures the salient aspects of the \OCaml implementation.
Our proof is mechanized in the \Coq proof assistant, using the \Iris base logic.
An independent contribution of our work is an extension of the \Simuliris approach to define simulation relations that support different calling conventions.
To our knowledge, this is the first use of \Simuliris to prove the correctness of a compiler transformation.

%% file: introduction.tex
\section{Introduction}

\subsection{Prologue}

``\OCaml'', we teach our students, ``is a functional programming language. We can write the beautiful function \ocaml{List.map} as follows:''
\begin{Ocaml}
let rec map f = function
| [] -> []
| x :: xs -> f x :: map f xs
\end{Ocaml}

``Well, actually, this version fails with a \ocaml{Stack_overflow}
exception on large input lists. If you want your \ocaml{map} to behave
correctly on all inputs, you should write a \emph{tail-recursive}
version. For this you can use the accumulator-passing style:''
\begin{Ocaml}
let map f li =
  let rec map_ acc = function
  | [] -> List.rev acc
  | x :: xs -> map_ (f x :: acc) xs
  in map_ [] f li
\end{Ocaml}

``Well, actually, this version works fine on large lists, but it is
less efficient than the original version. One approach is to start with
a non-tail-recursive version, and switch to a tail-recursive version
for large inputs; even there you can use some manual unrolling to
reduce the overhead of the accumulator. For example, the nice
\href{https://github.com/c-cube/ocaml-containers}{Containers} library
does it as follows:''.

\begin{minipage}{0.6\linewidth}
\begin{Ocaml}[basicstyle=\ttfamily\tiny]
let tail_map f l =
  (* Unwind the list of tuples, reconstructing the full list front-to-back.
     @param tail_acc a suffix of the final list; we append tuples' content
     at the front of it *)
  let rec rebuild tail_acc = function
    | [] -> tail_acc
    | (y0, y1, y2, y3, y4, y5, y6, y7, y8) :: bs ->
      rebuild (y0 :: y1 :: y2 :: y3 :: y4 :: y5 :: y6 :: y7 :: y8 :: tail_acc) bs
  in
  (* Create a compressed reverse-list representation using tuples
     @param tuple_acc a reverse list of chunks mapped with [f] *)
  let rec dive tuple_acc = function
    | x0 :: x1 :: x2 :: x3 :: x4 :: x5 :: x6 :: x7 :: x8 :: xs ->
      let y0 = f x0 in let y1 = f x1 in let y2 = f x2 in
      let y3 = f x3 in let y4 = f x4 in let y5 = f x5 in
      let y6 = f x6 in let y7 = f x7 in let y8 = f x8 in
      dive ((y0, y1, y2, y3, y4, y5, y6, y7, y8) :: tuple_acc) xs
    | xs ->
      (* Reverse direction, finishing off with a direct map *)
      let tail = List.map f xs in
      rebuild tail tuple_acc
  in
  dive [] l
\end{Ocaml}
\end{minipage}
\hfill
\begin{minipage}{0.4\linewidth}
\begin{Ocaml}[basicstyle=\ttfamily\tiny]
let direct_depth_default_ = 1000

let map f l =
  let rec direct f i l = match l with
    | [] -> []
    | [x] -> [f x]
    | [x1;x2] -> let y1 = f x1 in [y1; f x2]
    | [x1;x2;x3] ->
      let y1 = f x1 in let y2 = f x2 in [y1; y2; f x3]
    | _ when i=0 -> tail_map f l
    | x1::x2::x3::x4::l' ->
      let y1 = f x1 in
      let y2 = f x2 in
      let y3 = f x3 in
      let y4 = f x4 in
      y1 :: y2 :: y3 :: y4 :: direct f (i-1) l'
  in
  direct f direct_depth_default_ l
\end{Ocaml}
\end{minipage}
\lstset{basicstyle=\small\ttfamily}

At this point, unfortunately, some students leave the class and never
come back.

We propose a new feature for the \OCaml compiler, an explicit, opt-in
``Tail Modulo Cons'' transformation, to retain our students. After the
first version (or maybe, if we are teaching an advanced class, after
the second version), we could show them the following version:
\begin{Ocaml}
let[@tail_mod_cons] rec map f = function
| [] -> []
| x :: xs -> f x :: map f xs
\end{Ocaml}

This version is as fast as the simple implementation, tail-recursive,
and easy to write.

The catch, of course, is to teach when this \ocaml{[@tail_mod_cons]}
annotation can be used. Maybe we would not show it at all, and pretend
that the direct \ocaml{map} version with \ocaml{let y} is fine. This
would be a much smaller lie than it currently is,
a \ocaml{[@tail_mod_cons]}-sized lie.

Finally, experts should be very happy. They know about all these
versions, but they do not have to write them by hand anymore. Have
a program perform (some of) the program transformations that they are
currently doing manually.

\subsection{TMC transformation example}
\label{subsec:tmc_example}

A function call is in \emph{tail position} within a function
definition if the definition has ``nothing to do'' after evaluating
the function call -- the result of the call is the result of the whole
function at this point of the program. (A precise definition will be
given in Section~\ref{subsec:specification}.) A function is \emph{tail recursive}
if all its recursive calls are tail calls.

In the naive definition of \ocaml{map}, the recursive call is not in tail
position: after computing the result of \ocaml{map f xs} we still have
to compute the final list cell, \ocaml{y :: ?}. We say that a call is
\emph{tail modulo cons} when the remaining work is formed of data
\emph{constructors} only, such as \ocaml{(::)} here.

Other datatype constructors may be used; this is
also tail-recursive \emph{modulo cons}:

\begin{Ocaml}
let[@tail_mod_cons] rec tree_of_list = function
| [] -> Empty
| x :: xs -> Node(Empty, x, tree_of_list xs)
\end{Ocaml}

The TMC transformation produces an equivalent function in
\emph{destination-passing} style where the calls in \emph{tail modulo
  cons} position have been turned into \emph{tail} calls. In
particular, for \ocaml{map} it gives a tail-recursive function, which
runs in constant stack space; other list functions also become
tail-recursive. This works for other data types as well, such as binary trees,
but in this case some recursive calls may remain non-tail-recursive.

For \ocaml{map}, our transformation produces the following code:\\
\hspace{-1.6em}
\begin{minipage}{0.5\linewidth}
\begin{Ocaml}
let rec map f = function
| [] -> []
| x::xs ->
  let y = f x in
  let dst = y :: ? in
  map_dps dst 1 f xs;
  dst
\end{Ocaml}
\end{minipage}
\hfill
\begin{minipage}{0.5\linewidth}
\begin{Ocaml}
and map_dps dst i f = function
| [] ->
  dst.i <- []
| x::xs ->
  let y = f x in
  let dst' = y :: ? in
  dst.i <- dst';
  map_dps dst' 1 f xs
\end{Ocaml}
\end{minipage}

The transformed code has two variants of the \ocaml{map} function. The \ocaml{map_dps} variant is in \emph{destination-passing style}: it expects additional parameters that specify a memory location, a \emph{destination}, and writes its result to this \emph{destination} instead of returning it. It is tail-recursive, and it performs a single traversal of the list. The \ocaml{map} variant provides the same interface as the non-transformed function: we say that it is in \emph{direct style}. It is not tail-recursive, but it does not call itself recursively, it calls the tail-recursive \ocaml{map_dps} on non-empty lists.\footnote{The direct-style version of \ocaml{map} we produce is not recursive. But in the general case, the two functions produced may call each other, so we always produce a mutually-recursive block.}

The key idea of the transformation is that the expression \ocaml{y :: map f xs}, which contained a non-tail-recursive call, is transformed into: first create a \emph{partial} list cell, written \ocaml{y :: ?}, then call \ocaml{map_dps}, asking it to write its result in the position of the \ocaml{?} in the the partial cell. The recursive call thus takes place after the cell creation (instead of before), in tail-recursive position in the \ocaml{map_dps} variant. In the direct variant, the destination cell \ocaml{dst} is returned after the call.

The transformed code is pseudo-\OCaml: it is not a valid \OCaml
program. We use a magical \ocaml{?} constant, and our notation
\ocaml{dst.i <- ...} to update constructor parameters in-place is also
invalid in source programs. The transformation is implemented on
a lower-level, untyped intermediate representation of the \OCaml
compiler (Lambda), where those operations do exist. The \OCaml type
system is not expressive enough to type-check the transformed program:
the list cell is only partially initialized at first, each partial
cell is mutated exactly once, and in the end the whole result is
returned as an \emph{immutable} list. Some type systems are expressive
enough to represent this transformed code, notably \Mezzo~\citep*{mezzo-2016}, based on a permission system inspired by \emph{separation logic}~\citep*{seplog-cacm-2019}, or the linear types used in \citet*{minamide-98}.\Xfrancois{Voir aussi: Bagrel et Spiwack, JFLA 2024.}

TMC has been first implemented in Lisp~\citep*{risch-73,friedman-wise-75} and is well-known in the Lisp and Scheme implementation communities, but less well-known in other functional languages despite a few implementations\citep*{opal-1994,tmc-scala-2013}. A notable recent implementation (simultaneous with our work) is the one in Koka~\citep*{tmc-koka-2023}, which was carefully designed to support multishot delimited continuations. The TMC transform is arguably un-necessary in Prolog, where unification variables make it easy and idiomatic to express the transformed program, at the cost of a constant-factor overhead. A variant of TMC, which transforms recursive calls in tail-position modulo associative operations (rather than data constructors) into \emph{accumulator-passing style}, is in \texttt{gcc} and \texttt{clang}, allowing them to compile a naive definition of \ocaml{factorial} into a loop.

The first main contribution of our work is an implementation of TMC in the \OCaml compiler as an on-demand program transformation,
merged in November 2021.
We describe the non-trivial design choices in terms of user interface, and evaluate performance through micro-benchmarks.
Various functions in the standard library and third-party code bases have been rewritten to use it, to become tail-recursive, gain in performance, or (when an efficient but complex tail-recursive version was used) simplify considerably the implementation.

The second main contribution of this work is a mechanized proof of correctness for the core of this transformation on a small untyped calculus.
We establish that for any input source program there is a termination-preserving \emph{behavioral refinement} between the source program and the corresponding transformed program: any behavior of the transformed program, be it converging, diverging or stuck, is a behavior of the source program.
To our knowledge this is the first verification of the TMC transformation in an untyped setting.

Our proof technique is to define a relational program logic for our small untyped calculus, to show the correctness of the TMC transformation using this program logic, and to get the behavioral refinement by proving adequacy of our program logic. We build on top of \Simuliris~\citep*{simuliris-2022}, a framework for simulations in separation logic over the \Iris base logic. The use of separation logic nicely captures certain aspects of the proof argument, in particular the fact that the the destination-passing-style function uniquely owns the destination location that it receives.

To our knowledge, previous works on \Simuliris have verified \emph{examples} of interesting optimizations and program transformations, by formally proving relations between pairs of concrete programs. Our work may be the first proof of correctness of a \emph{program transformation} (as a function or relation) using a simulation-based approach, establishing correctness for all input programs. (Earlier \Iris work proves program transformations using logical relations, see for example \citet*{tassarotti-2017}.)

At this level of generality, we found that the \Simuliris simulation is not expressive enough to reason about transformations that introduce new function calling conventions.
We generalize the \Simuliris handling of function calls by parameterizing the simulation relation over an \emph{abstract protocol}, inspired by \citet*{protocols-2021}.
This sub-contribution of our work is independent from the TMC transformation and our small calculus,
and we tried to express it in general terms, beyond the specific needs of TMC.
In particular, we believe that \Iris-based relational separation logics could be a powerful yet pleasant proof technique for compiler verification.

The core of the soundness proof is the specification of the two variants of each TMC-transformed function --- direct style and destination-passing style.
It concisely conveys the essence of destination-passing style: computing the same thing and writing it to an owned destination.
For instance, to give a taste of the formalism, the specification of the variants of \ocaml|map| looks as follows:
\begin{align*}
        \iSimvHoare{
            \datalangVal_s \iSimilar \datalangVal_t
        }{
            \iSimilar
        }{
          & \datalangCall
              {\datalangFnptr{\ocamlText{map}}}
              {\datalangPair{\datalangFnptr{\datalangFn}}{\datalangVal_s}}
        }{
            \datalangCall
              {\datalangFnptr{\ocamlText{map}}}
              {\datalangPair{\datalangFnptr{\datalangFn}}{\datalangVal_t}}
        }
    \\
        \iSimvHoare{
            \datalangVal_s \iSimilar \datalangVal_t
            \ \iSep\ %
            (\datalangLoc + \datalangIdx) \iPointsto \datalangHole
        }{
            \datalangVal'_s, \datalangUnit \ldotp
            \exists \datalangVal'_t \ldotp\ %
            (\datalangLoc + \datalangIdx) \iPointsto \datalangVal'_t
            \ \iSep\ %
            \datalangVal'_s \iSimilar \datalangVal'_t
        }{
          & \datalangCall
              {\datalangFnptr{\ocamlText{map}}}
              {\datalangPair{\datalangFnptr{\datalangFn}}{\datalangVal_s}}
        }{
            \datalangCall
              {\datalangFnptr{\ocamlText{map_dps}}}
              {\datalangTriple
                {\datalangPair{\datalangLoc}{\datalangIdx}}
                {\datalangFnptr{\datalangFn}}
                {\datalangVal_t}}
        }
\end{align*}
If two input lists are related, then calling the \ocaml{map} function or its direct-style translation will return related outputs. Furthermore, if we call the destination-passing-style version on a partial block that we own, we will get a source value $\datalangVal'_s$ and a \ocaml{unit} value $\datalangUnit$, and a target value $\datalangVal'_t$ related to $\datalangVal'_s$ will be written in the block.

To sum up, our main contributions are:
\begin{enumerate}
    \item an implementation of the TMC transformation in the \OCaml compiler, with a discussion of the user interface, a performance evaluation, and a survey of its early usage;
    \item a mechanized proof of soundness for an idealized TMC transformation on a small calculus, using a relational separation program logic;
    \item a generalization of the \Simuliris handling of function calls with abstract \emph{protocols} to reason about different calling conventions.
\end{enumerate}

\paragraph{Remarks}
A preliminary, work-in-progress version of this work was presented in
a previous publication at a national conference~\citep*{tmc-ocaml-2021}.
\begin{version}{\FALSE}
  We explain our choice to submit a full paper to an international
  conference in \cref{app:extended-republication}.
\end{version}
Our mechanized proofs are available at \url{https://zenodo.org/records/13937565}.


%% file: formalization.tex
\section{TMC on an idealized language}
\label{sec:formalization}

In this section, we formalize the ``tail modulo cons'' (TMC) transformation in an idealized language, \DataLang, that is expressive enough to account for the main aspects of TMC but does not support all features of \OCaml.
Our proof of correctness covers this idealized fragment.
We intentionally keep the presentation very close to our Coq development, which can be referred to for full details.

\input{figures/syntax}
\input{figures/sugar}
\input{figures/semantics}
\input{figures/map}

\subsection{Language}

The syntax of \DataLang is given in \cref{fig:syntax} and its semantics in \cref{fig:semantics}.
We also introduce syntactic sugar in \cref{fig:sugar}, in particular shallow pattern-matching on lists.
Going back to our motivating example, we can define the \datalang|map| function on lists as in \cref{fig:map}.

\DataLang is an untyped sequential calculus with mutable state. A \DataLang program $\datalangProg$ is a finite mapping from function names $\datalangFn \in \datalangFn[]$ to mutually-recursive definitions $d$, which are themselves functions whose body is written $\datalangRec \datalangVar \datalangExpr$. Functions have a single parameter for simplicity, with pairs used to pass several values.

\DataLang has Booleans $b \in \{\datalangTrue, \datalangFalse\}$, an if-then-else construct, and a runtime equality test\footnote{We check physical equality on locations / pointers, and primitive equality between primitive types, similarly to the \texttt{eqv?} predicate of Scheme. Primitive values of distinct types, for example integers and Booleans, are always considered different.} between (untyped) values $\datalangEq{\datalangExpr_1}{\datalangExpr_2}$.

\DataLang is first-order in the same sense that C is (with function pointers): it does not feature general lambda expressions, its programs correspond to closure-converted or lambda-lifted source programs.\footnote{
The usual definition of TMC that we implement and formalize is essentially first-order.%
 See \Appendices{\cref{subsubsec:first-order}}{Appendix~A.2.3}.}
Functions names $\datalangFn$ can be turned into values written $\datalangFnptr{\datalangFn}$, to be used directly in function calls or as parameters to higher-order functions.

To express constructors, \DataLang features mutable memory blocks with an abstract \emph{tag} ($\datalangTag \in \datalangTag[]$), and two \emph{fields} which are arbitrary values ($\datalangExpr_1, \datalangExpr_2$).
One can allocate a block with $\datalangBlock{\datalangTag}{\datalangExpr_1}{\datalangExpr_2}$, access its fields with $\datalangLoad{\datalangExpr_1}{\datalangExpr_2}$ and modify them with $\datalangStore{\datalangExpr_1}{\datalangExpr_2}{\datalangExpr_3}$.
Allocation returns a location $\datalangLoc \in \datalangLoc[]$, which may not appear in source programs.

The evaluation order of subexpressions $\datalangExpr_1$ and $\datalangExpr_2$ in $\datalangBlock{\datalangTag}{\datalangExpr_1}{\datalangExpr_2}$ is unspecified as in \OCaml.
This is crucial to allow the behavior-preserving optimization of more programs, as the TMC transformation may affect the evaluation order of subterms of data constructors.
To model this in the semantics, we introduce a separate, deterministic block construction $\datalangBlockDet{\datalangTag}{\datalangExpr_1}{\datalangExpr_2}$ which cannot appear in source programs.
A block expression $\datalangBlock{\datalangTag}{\datalangExpr_1}{\datalangExpr_2}$
first reduces (in a nondeterministic manner)
to either $\datalangLet{\datalangVar_1}{\datalangExpr_1}{\datalangLet{\datalangVar_2}{\datalangExpr_2}{\datalangBlockDet{\datalangTag}{\datalangVar_1}{\datalangVar_2}}}$ through \RefTirName{StepBlock1}
or to $\datalangLet{\datalangVar_2}{\datalangExpr_2}{\datalangLet{\datalangVar_1}{\datalangExpr_1}{\datalangBlockDet{\datalangTag}{\datalangVar_1}{\datalangVar_2}}}$ through \RefTirName{StepBlock2}.
$\datalangBlockDet{\datalangTag}{\datalangVal_1}{\datalangVal_2}$ performs the allocation through \RefTirName{StepBlockDet}.

The values in \DataLang are functions $\datalangFnptr{\datalangFn}$, locations $\datalangLoc$ of allocated blocks, Booleans $b$, tags $t$ (taken in an arbitrary, denumerable set), the unit value $\datalangUnit$, and indices $i \in \{\datalangZero, \datalangOne, \datalangTwo\}$ inside blocks. (Our transformation never mutates block tags, nor do \OCaml programs, but \Mezzo supports it.)

On top of these basic language features, \cref{fig:sugar} introduces syntactic sugar for pairs $\datalangPair {\datalangExpr_1} {\datalangExpr_2}$ as blocks with a specific tag $\mathrm{PAIR}$, for decomposing blocks in $\term{\color{datalangKeyword1}{let}}$-bindings ($\datalangLet {\datalangPair \datalangVar \datalangVarTwo} {\datalangExpr_1} {\datalangExpr_2}$) and in arguments of toplevel functions ($\datalangFn \mapsto \datalangRec {\datalangPair \datalangVar \datalangVarTwo} \datalangExpr$), and for (untyped) lists by defining the empty list $\datalangNil$ as $\datalangUnit$, and for (mutable) cons-cells as blocks with a specific tag $\mathrm{CONS}$. Pattern-matching on lists can be expressed by comparing the list with $\datalangNil$, using our block-deconstructing $\term{\color{datalangKeyword1}{let}}$ (which ignores the tag) to deconstruct cons-cells.

As a side note, we use named expression variables here but the \Coq mechanization actually adopts de Bruijn syntax, which is better suited to define transformations involving binders.
More precisely, our formalization relies on the \Autosubst library~\cite{autosubst-2015}.
Our definitions respect $\alpha$-equivalence on term variables $\datalangVar, \datalangVarTwo$: we implicitly assume any term variable in bound position to be chosen distinct from all other variables in context.
Function names $\datalangFn$ are not $\alpha$-renamed, as the transformation relates names in the source and target of the transformation.

\subsection{Transformation}
\label{subsec:transformation}

We now define the TMC transformation as a relation $\datalangProg_s \tmc \datalangProg_t$ between programs and their transformation.
The relation is total, in the sense that any \DataLang program $\datalangProg_s$ can be related to at least one transformed program $\datalangProg_t$.
It is not deterministic: for each input program it captures a (finite) set of admissible transformations, which we all prove valid.
This non-determinism captures several choices that have to be done by the user through a user interface to control the transformation, or by the compiler implementation, influencing performance and evaluation order of the result.
In this section, we do not describe how these choices are resolved -- there is a large design space.
We present the choices we made for \OCaml compiler in \cref{sec:implementation}.

As formalized in \cref{fig:tmc}, transforming a \DataLang program $\datalangProg$ consists in:

\input{figures/tmc}
\paragraph{1. Choosing a subset of toplevel functions to be TMC-transformed}
For each such function $\datalangFn$, we also require a fresh function name $\datalangRenaming [\datalangFn]$ (that is not defined in $\datalangProg_s$) that will be the \emph{destination-passing style} (DPS) version of $\datalangFn$ in the transformed program $\datalangProg_t$.

Formally, the subset is determined by the domain of the renaming function $\datalangRenaming$, which is passed as a parameter to the auxiliary transformations that we describe next.

\input{figures/tmc_dir}
\paragraph{2. For each function $\datalangFn$ defined in $\datalangProg$, computing its direct transform.}
We introduce in \cref{fig:tmc_dir} the relations $\datalangDef_s \tmcDir{\datalangRenaming} \datalangDef_t$ for definitions and $\datalangExpr_s \tmcDir{\datalangRenaming} \datalangExpr_t$ for expressions.
$\datalangDef_s \tmcDir{\datalangRenaming} \datalangDef_t$ expresses that:
1)~$\datalangDef_t$ has the same calling convention as $\datalangDef_s$.
2)~The body of $\datalangDef_t$ is the direct transform of the body of $\datalangDef_s$.
$\datalangExpr_s \tmcDir{\datalangRenaming} \datalangExpr_t$ expresses that $\datalangExpr_t$ is the direct transform of $\datalangExpr_s$.
Intuitively: $\datalangExpr_t$ computes the same thing as $\datalangExpr_s$.

The direct-style transform corresponds to the case where we do not have a block that can serve as a destination: this version is used in an arbitrary calling context, not necessarily under a constructor. Most rules are straightforward congruences -- we recursively transform subexpressions and preserve the term constructor. We omit the rules for loads $\datalangLoad{\datalangExpr_1}{\datalangExpr_2}$, stores $\datalangStore{\datalangExpr_1}{\datalangExpr_2}{\datalangExpr_3}$, and the deterministic blocks $\datalangBlockDet \datalangTag {\datalangExpr_1} {\datalangExpr_2}$ which are such simple congruences, just like calls $\datalangCall {\datalangExpr_1} {\datalangExpr_2}$.

The key cases are for a block construct $\datalangBlock \datalangTag {\datalangExpr_1} {\datalangExpr_2}$. We can use this block as a destination, and switch to the destination-passing-style calling convention -- these rules are a source of non-determinism, and the only places in the direct-style transformation where destination-passing-style is introduced. \RefTirName{DirBlock} is a simple congruence rule that keeps both arguments in direct style. The rules (\RefTirName{DirBlockDPS1}, \RefTirName{DirBlockDPS2}) choose a block argument to be evaluated in destination-passing style. (It is also possible to transform both arguments in DPS style, and we include extra rules for this in our formalization. \Xgabriel{TODO: measure the performance of this tweak for \ocaml{map} on binary trees.})

The terms produced by these rules proceed as follows:
1)~Partially initialize a new memory block, with a hole for one of their arguments.
2)~Evaluate the DPS transformation of the corresponding argument,
   passing the uninitialized field as destination.
3)~Return the now fully initialized block.

An implementation would typically determine which subexpression would benefit from destina\-tion-passing style, that is, contains function calls $\datalangCall {\datalangFnptr \datalangFn} \datalangExpr$ in tail position (relatively to the subexpression) that have a destination-passing variant $\datalangRenaming [\datalangFn]$.

\input{figures/tmc_dps}
\paragraph{3. For each TMC-transformed function $\datalangFn$, choosing a destination-passing-style transform}
We introduce in \cref{fig:tmc_dps} the relations $\datalangDef_s \tmcDps{\datalangRenaming} \datalangDef_t$ for definitions and $(\datalangExpr_\mathit{dst}, \datalangExpr_\mathit{idx}, \datalangExpr_s) \tmcDps{\datalangRenaming} \datalangExpr_t$ for expressions.

$\datalangDef_s \tmcDps{\datalangRenaming} \datalangDef_t$ expresses that:
1)~The function defined in $\datalangDef_t$ has an additional parameter representing the destination where it must write its result. This parameter is a pair of the location of a memory block $\datalangVar_\mathit{dst}$ along with the index $\datalangVar_\mathit{idx}$ of a particular field in this block.
2)~The body of $\datalangDef_t$ is a DPS transform of the body of $\datalangDef_s$ under the given destination.

$(\datalangExpr_\mathit{dst}, \datalangExpr_\mathit{idx}, \datalangExpr_s) \tmcDps{\datalangRenaming} \datalangExpr_t$ expresses that $\datalangExpr_t$ is a DPS transform of $\datalangExpr_s$ under destination $(\datalangExpr_\mathit{dst}, \datalangExpr_\mathit{idx})$.
Intuitively, this means $\datalangExpr_t$ computes the same thing as $\datalangExpr_s$ but writes it into the destination instead of returning it.
We will formalize this intuition in \cref{sec:specification}.

Note that the rule \RefTirName{DPSDef}, which relates the two judgments, uses the expression-level relation $(\mathrel{\tmcDps{\datalangRenaming}})$ with a term variable $\datalangVar_\mathit{idx}$ to represent the index, not just a constant index $\datalangOne$ or $\datalangTwo$ as in block rules. These are the only two sort of expressions used to represent offsets in the transformation.

In the direct-style relation, congruence rules apply the same direct-style transformation to all subexpressions. The congruence-like rule of the DPS relation, for example \RefTirName{DPSLet} and \RefTirName{DPSIf}, are different. They apply the DPS transformation to sub-expressions which are in tail position relative to the expression, and the direct-style transformation to all other subexpressions. The if-then-else construct has two different subexpressions in tail position, only one of them is evaluated at runtime.

The rules \RefTirName{DPSBlock1} and \RefTirName{DPSBlock2}
correspond to the rules \RefTirName{DirBlockDPS1} and \RefTirName{DirBlockDPS2} in the direct-style transformation, but the transformed code is different. Consider the translation of $
\left(
  \datalangExpr_\mathit{dst},
  \datalangExpr_\mathit{idx},
  \datalangBlock{\datalangTag}{\datalangExpr_{s1}}{\datalangExpr_{s2}}
\right)
$ into $
\datalangLet{\datalangVar}{\datalangBlock{\datalangTag}{\datalangExpr_{t1}}{\datalangHole}}{
  \datalangSeq{\datalangStore{\datalangExpr_\mathit{dst}}{\datalangExpr_\mathit{idx}}{\datalangVar}}{\datalangExpr_{t2}}}
$ by \RefTirName{DirBlockDPS2}. First we create a new destination $\datalangVar$, with a hole in second position. Then, instead of computing the corresponding subterm, we write this new destination $x$ into the \emph{current} destination $({\datalangExpr_\mathit{dst}}, {\datalangExpr_\mathit{idx}})$. Finally we evaluate $\datalangExpr_{t2}$, which is the DPS transform of the subexpression $\datalangExpr_{s2}$, with the destination $(\datalangVar, \datalangTwo)$. Notice that $\datalangExpr_{t2}$ is in tail position relative to the transformed expression, while it was not in tail position in the source expression. This is the key step of the TMC transformation, that turns non-tail calls into tail calls. Rule \RefTirName{DirBlockDPS2} puts the second subterm in tail position, and \RefTirName{DirBlockDPS1} puts the first subterm in tail position. It is not always obvious which rule should be applied. In the case of lists as in our running example \ocaml|y :: map f xs|, we want the second subterm in tail position, so the transformation only uses \RefTirName{DirBlockDPS2}. But consider a \ocaml|map| function on binary trees \ocaml|Node(map f left, map f right)|: the implementation must choose one subterm to put in tail position and another to keep in non-tail position.

The rule \RefTirName{DPSCall} applies only to calls $\datalangCall {\datalangFnptr \datalangFn} {\datalangExpr_s}$ to a known function $\datalangFn$, on the condition that a DPS variant has been generated for $\datalangFn$: $\datalangFn \in \dom{\datalangRenaming}$. In this case, the function call can be compiled to a call to the DPS variant $\datalangRenaming [\datalangFn]$, transferring to the callee the responsibility to write to the destination. This is the case where the DPS transform is beneficial, as this transformation may turn a non-tail-call into a tail-call -- when it occurs under a block, in a subterm that was moved to tail position. This rule is selected for \ocaml|map| in our \ocaml|y :: map f xs| example.

Finally, there is a catch-all rule \RefTirName{DPSBase} that applies in any case, in particular whenever none of the other rules can be selected. This case trivially realizes the DPS calling convention by evaluating the subterm to a result and writing this result in the desired destination. This is what happens in the base case of \ocaml|map_dps|, where the empty list \ocaml|[]| is transformed into \ocaml|dst.(idx) <- []|.



\subsection{Realizing the relation as a function}

Our Coq formalization includes a function that takes an input program and outputs a related program, following the one-pass implementation approach that we introduced in the \OCaml compiler\Appendices{ (\cref{subsec:implementation})}{(see Appendix~A.4)}.


%% file: figures/syntax.tex
\begin{figure}[tp]
    \begin{tabular}{lclcl}
            $\datalangIdx[]$
            & $\ni$ &
            $\datalangIdx$
            & $\Coloneqq$ &
            $\datalangZero \mid \datalangOne \mid \datalangTwo$
        \\
            $\datalangBool[]$
            & $\ni$ &
            $\datalangBool$
            & $\Coloneqq$ &
            $\datalangTrue \mid \datalangFalse$
        \\
            $\datalangTag[]$
            & $\ni$ &
            $\datalangTag$
    	\\
    		$\datalangLoc[]$
    		& $\ni$ &
    		$\datalangLoc$
        \\
            $\datalangFn[]$
            & $\ni$ &
            $\datalangFn$
        \\
            $\datalangVar[]$
            & $\ni$ &
            $\datalangVar, \datalangVarTwo$
    	\\
            $\datalangVal[]$
            & $\ni$ &
            $\datalangVal, \datalangValTwo$
            & $\Coloneqq$ &
            $\datalangUnit \mid \datalangIdx \mid \datalangTag \mid \datalangBool \mid \datalangLoc \mid \datalangFnptr{\datalangFn}$
      \\
  \end{tabular}
  \begin{tabular}{lclcl}
            $\datalangDef[]$
            & $\ni$ &
            $\datalangDef$
            & $\Coloneqq$ &
            $\datalangRec{\datalangVar}{\datalangExpr}$
        \\
            $\datalangProg[]$
            & $\ni$ &
            $\datalangProg$
            & $\coloneqq$ &
            $\datalangFn[] \finmap \datalangDef[]$
        \\
            $\datalangState[]$
            & $\ni$ &
            $\datalangState$    
            & $\coloneqq$ &
            $\datalangLoc[] \finmap \datalangVal[]$
        \\
            $\datalangConfig[]$
            & $\ni$ &
            $\datalangConfig$
            & $\coloneqq$ &
            $\datalangExpr[] \times \datalangState[]$
    \end{tabular}

  \begin{tabular}{rcl}
            $\datalangExpr[]$
            $\ni$
            $\datalangExpr$
            & $\Coloneqq$ &
            $\datalangVal$
        \\
            & | &
            $\datalangVar \mid \datalangLet{\datalangVar}{\datalangExpr_1}{\datalangExpr_2} \mid \datalangCall{\datalangExpr_1}{\datalangExpr_2}$
        \\
            & | &
            $\datalangEq{\datalangExpr_1}{\datalangExpr_2}$
        \\
            & | &
            $\datalangIf{\datalangExpr_0}{\datalangExpr_1}{\datalangExpr_2}$
        \\
            & | &
            $\datalangBlock{\datalangTag}{\datalangExpr_1}{\datalangExpr_2} \mid \datalangBlockDet{\datalangTag}{\datalangExpr_1}{\datalangExpr_2}$
        \\
            & | &
            $\datalangLoad{\datalangExpr_1}{\datalangExpr_2} \mid \datalangStore{\datalangExpr_1}{\datalangExpr_2}{\datalangExpr_3}$
  \end{tabular}~
  \begin{tabular}{lclcl}
            $\datalangEctx[]$
            $\ni$
            $\datalangEctx$
            & $\Coloneqq$ &
            $\Box$
        \\
            & | &
            $\datalangLet{\datalangVar}{\datalangEctx}{\datalangExpr_2} \mid \datalangCall{\datalangExpr_1}{\datalangEctx} \mid \datalangCall{\datalangEctx}{\datalangVal_2}$
        \\
            & | &
            $\datalangEq{\datalangExpr_1}{\datalangEctx} \mid \datalangEq{\datalangEctx}{\datalangVal_2}$
        \\
            & | &
            $\datalangIf{\datalangEctx}{\datalangExpr_1}{\datalangExpr_2}$
        \\
            & | &
            $\datalangLoad{\datalangExpr_1}{\datalangEctx} \mid \datalangLoad{\datalangEctx}{\datalangVal_2}$
        \\
            & | &
            $\datalangStore{\datalangExpr_1}{\datalangExpr_2}{\datalangEctx} \mid \datalangStore{\datalangExpr_1}{\datalangEctx}{\datalangVal_3} \mid \datalangStore{\datalangEctx}{\datalangVal_2}{\datalangVal_3}$
  \end{tabular}
    \caption{\DataLang syntax}
    \label{fig:syntax}
\end{figure}


%% file: figures/sugar.tex
\begin{figure}[tp]
    \begin{tabular}{rcl}
            $\datalangSeq{\datalangExpr_1}{\datalangExpr_2}$
            & $\coloneqq$ &
            $\datalangLet{\datalangVar}{\datalangExpr_1}{\datalangExpr_2}$
        \\
            $\datalangPair{\datalangExpr_1}{\datalangExpr_2}$
            & $\coloneqq$ &
            $\datalangBlock{\mathrm{PAIR}}{\datalangExpr_1}{\datalangExpr_2}$
        \\
            $\datalangLet{\datalangPair{\datalangVar_1}{\datalangVar_2}}{\datalangExpr_1}{\datalangExpr_2}$
            & $\coloneqq$ &
            $\datalangLet{\datalangVarTwo}{\datalangExpr_1}{$
        \\
            &&
            $\datalangLet{\datalangVar_1}{\datalangLoad{\datalangVarTwo}{1}}{$
        \\
            &&
            $\datalangLet{\datalangVar_2}{\datalangLoad{\datalangVarTwo}{2}}{$
        \\
            &&
            $\datalangExpr_2}}}$
    \end{tabular}~%
    \begin{tabular}{rcl}
            $\datalangRec{\datalangPair{\datalangVar_1}{\datalangVar_2}}{\datalangExpr}$
            & $\coloneqq$ &
            $\datalangRec{\datalangVarTwo}{
            \datalangLet{\datalangPair{\datalangVar_1}{\datalangVar_2}}{\datalangVarTwo}{\datalangExpr}}$
        \\
            $\datalangNil$
            & $\coloneqq$ &
            $\datalangUnit$
        \\
            $\datalangCons{\datalangExpr_1}{\datalangExpr_2}$
            & $\coloneqq$ &
            $\datalangBlock{\mathrm{CONS}}{\datalangExpr_1}{\datalangExpr_2}$
    \end{tabular}

    \begin{tabular}{rcl@{}l}
            $\datalangMatchOneline[]{\datalangExpr_0}{\datalangExpr_1}{\datalangVar}{\mathit{\datalangVar s}}{\datalangExpr_2}$
            & $\coloneqq$ &
            $\datalangLet{\datalangVarTwo}{\datalangExpr_0}{
               \datalangIf{\datalangEq{\datalangVarTwo}{\datalangNil}$&$}{\datalangExpr_1$ 
        \\
            &&&
            $}{\datalangLet{\datalangPair{\datalangVar}{\mathit{\datalangVar s}}}{\datalangVarTwo}{\datalangExpr_2}}}$
        \\
            $\datalangHole$
            & $\coloneqq$ &
            $\datalangUnit$ &
    \end{tabular}
    \caption{\DataLang syntactic sugar}
    \label{fig:sugar}
\end{figure}

%% file: figures/semantics.tex
\begin{figure}[tp]
    \begin{mathparpagebreakable}
        \inferrule*
            {}{
                \boxed{- \headStep{\datalangProg} - : \datalangConfig[] \rightarrow \datalangConfig[] \rightarrow \Prop}
            }
        \and
        \inferrule*
            {}{
                \boxed{- \step{\datalangProg} - : \datalangConfig[] \rightarrow \datalangConfig[] \rightarrow \Prop}
            }
        \\
        \inferrule*[lab=StepLet]
            {}{
                \left(
                    \datalangLet{\datalangVar}{\datalangVal}{\datalangExpr},
                    \datalangState
                \right)
                \headStep{\datalangProg}
                \left(
                    \datalangExpr{} [\datalangVar \backslash \datalangVal],
                    \datalangState
                \right)
            }
        \and
        \inferrule*[lab=StepCall]
            {
                \datalangProg{} [\datalangFn] = (\datalangRec{\datalangVar}{\datalangExpr})
            }{
                \left(
                    \datalangCall{\datalangFnptr{\datalangFn}}{\datalangVal},
                    \datalangState
                \right)
                \headStep{\datalangProg}
                \left(
                    \datalangExpr{} [\datalangVar \backslash \datalangVal],
                    \datalangState
                \right)
            }
        \\
        \inferrule*[lab=StepBlock1]
            {}{
                \left(
                    \datalangBlock{\datalangTag}{\datalangExpr_1}{\datalangExpr_2},
                    \datalangState
                \right)
                \headStep{\datalangProg}
                \left(
                    \begin{array}{l}
                        \datalangLet{\datalangVar_1}{\datalangExpr_1}{\\
                        \datalangLet{\datalangVar_2}{\datalangExpr_2}{\\
                        \datalangBlockDet{\datalangTag}{\datalangVar_1}{\datalangVar_2}}}
                    \end{array},
                    \datalangState
                \right)
            }
        \and
        \inferrule*[lab=StepBlock2]
            {}{
                \left(
                    \datalangBlock{\datalangTag}{\datalangExpr_1}{\datalangExpr_2},
                    \datalangState
                \right)
                \headStep{\datalangProg}
                \left(
                    \begin{array}{l}
                        \datalangLet{\datalangVar_2}{\datalangExpr_2}{\\
                        \datalangLet{\datalangVar_1}{\datalangExpr_1}{\\
                        \datalangBlockDet{\datalangTag}{\datalangVar_1}{\datalangVar_2}}}
                    \end{array},
                    \datalangState
                \right)
            }
        \\
        \inferrule*[lab=StepBlockDet]
            {
                \forall \datalangIdx \in \datalangIdx[], \datalangLoc + \datalangIdx \notin \dom{\datalangState}
            }{
                \left(
                    \datalangBlockDet{\datalangTag}{\datalangVal_1}{\datalangVal_2},
                    \datalangState
                \right)
                \headStep{\datalangProg}
                \left(
                    \datalangLoc,
                    \datalangState{} [\datalangLoc \mapsto \datalangTag, \datalangVal_1, \datalangVal_2]
                \right)
            }
        \and
        \inferrule*[lab=StepLoad]
            {
                \datalangState{} [\datalangLoc + \datalangIdx] = \datalangVal
            }{
                \left(
                    \datalangLoad{\datalangLoc}{\datalangIdx},
                    \datalangState
                \right)
                \headStep{\datalangProg}
                \left(
                    \datalangVal,
                    \datalangState
                \right)
            }
        \and
        \inferrule*[lab=StepStore]
            {
                \datalangLoc + \datalangIdx \in \dom{\datalangState}
            }{
                \left(
                    \datalangStore{\datalangLoc}{\datalangIdx}{\datalangVal},
                    \datalangState
                \right)
                \headStep{\datalangProg}
                \left(
                    \datalangUnit,
                    \datalangState{} [\datalangLoc + \datalangIdx \mapsto \datalangVal]
                \right)
            }
        \and
        \inferrule*[lab=StepEctx]
            {
                \left(
                    \datalangExpr,
                    \datalangState
                \right)
                \headStep{\datalangProg}
                \left(
                    \datalangExpr',
                    \datalangState'
                \right)
            }{
                \left(
                    \datalangEctx{} [\datalangExpr],
                    \datalangState
                \right)
                \step{\datalangProg}
                \left(
                    \datalangEctx{} [\datalangExpr'],
                    \datalangState'
                \right)
            }
    \end{mathparpagebreakable}
    \caption{\DataLang semantics (excerpt)}
    \label{fig:semantics}
\end{figure}

%% file: figures/map.tex
\begin{figure}[tp]
\begin{tabular}{c}
\begin{Datalang}
map |-> rec λ(f, xs) = match xs with
                       | [] -> []
                       | x :: xs -> let y = f x in y :: @map (f, xs)
\end{Datalang}
\end{tabular}
\caption{Natural implementation of \datalang|map| in \DataLang}
\label{fig:map}
\end{figure}

%% file: figures/tmc.tex
\begin{figure}[tp]
    \[
        \datalangProg_s \tmc \datalangProg_t
        \coloneqq
        \exists \datalangRenaming \ldotp
        \bigwedge \left[ \begin{array}{l}
                \dom{\datalangRenaming} \subseteq \dom{\datalangProg_s}
            \\
                \dom{\datalangProg_t} = \dom{\datalangProg_s} \cup \codom{\datalangRenaming}
            \\
                \forall \datalangFn \in \dom{\datalangProg_s} \ldotp \quad
                \datalangProg_s[\datalangFn] \tmcDir{\datalangRenaming} \datalangProg_t[\datalangFn]
            \\
                \forall (\datalangFn \mapsto \datalangFn_\mathit{dps}) \in \datalangRenaming, \quad
                \datalangProg_s[\datalangFn] \tmcDps{\datalangRenaming} \datalangProg_t[\datalangFn_\mathit{dps}]
        \end{array} \right.
    \]
    \caption{TMC transformation}
    \label{fig:tmc}
\end{figure}

%% file: figures/tmc_dir.tex
\begin{figure}[tp]
    \begin{mathparpagebreakable}
        \inferrule*
            {}{
                \boxed{- \tmcDir{\datalangRenaming} - : \datalangDef[] \rightarrow \datalangDef[] \rightarrow \Prop}
            }
        \and
        \inferrule*
            {}{
                \boxed{- \tmcDir{\datalangRenaming} - : \datalangExpr[] \rightarrow \datalangExpr[] \rightarrow \Prop}
            }
        \and
        \inferrule*[lab=DirDef]
            {
                \datalangExpr_s \tmcDir{\datalangRenaming} \datalangExpr_t
            }{
                \datalangRec{\datalangVar}{\datalangExpr_s}
                \tmcDir{\datalangRenaming}
                \datalangRec{\datalangVar}{\datalangExpr_t}
            }
        \and
        \inferrule*[lab=DirVal]
            {}{
                \datalangVal
                \tmcDir{\datalangRenaming}
                \datalangVal
            }
        \and
        \inferrule*[lab=DirVar]
            {}{
                \datalangVar
                \tmcDir{\datalangRenaming}
                \datalangVar
            }
        \and
        \inferrule*[lab=DirLet]
            {
                \datalangExpr_{s1} \tmcDir{\datalangRenaming} \datalangExpr_{t1}
            \and
                \datalangExpr_{s2} \tmcDir{\datalangRenaming} \datalangExpr_{t2}
            }{
                \datalangLet{\datalangVar}{\datalangExpr_{s1}}{\datalangExpr_{s2}}
                \tmcDir{\datalangRenaming}
                \datalangLet{\datalangVar}{\datalangExpr_{t1}}{\datalangExpr_{t2}}
            }
        \and
        \inferrule*[lab=DirCall]
            {
                \datalangExpr_{s1} \tmcDir{\datalangRenaming} \datalangExpr_{t1}
            \and
                \datalangExpr_{s2} \tmcDir{\datalangRenaming} \datalangExpr_{t2}
            }{
                \datalangCall {\datalangExpr_{s1}} {\datalangExpr_{s2}}
                \tmcDir{\datalangRenaming}
                \datalangCall {\datalangExpr_{t1}} {\datalangExpr_{t2}}
            }
        \and
        \inferrule*[lab=DirBlock]
            {
                \datalangExpr_{s1} \tmcDir{\datalangRenaming} \datalangExpr_{t1}
            \and
                \datalangExpr_{s2} \tmcDir{\datalangRenaming} \datalangExpr_{t2}
            }{
                \datalangBlock \datalangTag {\datalangExpr_{s1}} {\datalangExpr_{s2}}
                \tmcDir{\datalangRenaming}
                \datalangBlock \datalangTag {\datalangExpr_{t1}} {\datalangExpr_{t2}}
            }
        \and
        \inferrule*[lab=DirBlockDPS1]
            {
                (\datalangVar, \datalangOne, \datalangExpr_{s1}) \tmcDps{\datalangRenaming} \datalangExpr_{t1}
            \and
                \datalangExpr_{s2} \tmcDir{\datalangRenaming} \datalangExpr_{t2}
            }{
                \datalangBlock{\datalangTag}{\datalangExpr_{s1}}{\datalangExpr_{s2}}
                \tmcDir{\datalangRenaming}
                \begin{array}{l}
                    \datalangLet{\datalangVar}{\datalangBlock{\datalangTag}{\datalangHole}{\datalangExpr_{t2}}}{\\
                    \datalangSeq{\datalangExpr_{t1}}{\datalangVar}}
                \end{array}
            }
        \and
        \inferrule*[lab=DirBlockDPS2]
            {
                \datalangExpr_{s1} \tmcDir{\datalangRenaming} \datalangExpr_{t1}
            \and
                (\datalangVar, \datalangTwo, \datalangExpr_{s2}) \tmcDps{\datalangRenaming} \datalangExpr_{t2}
            }{
                \datalangBlock{\datalangTag}{\datalangExpr_{s1}}{\datalangExpr_{s2}}
                \tmcDir{\datalangRenaming}
                \begin{array}{l}
                    \datalangLet{\datalangVar}{\datalangBlock{\datalangTag}{\datalangExpr_{t1}}{\datalangHole}}{\\
                  \datalangSeq{\datalangExpr_{t2}}{\datalangVar}}
                \end{array}
            }
    \end{mathparpagebreakable}
    \caption{Direct TMC transformation (omitting congruence rules similar to \TirName{DirCall})}
    \label{fig:tmc_dir}
\end{figure}


%% file: figures/tmc_dps.tex
\begin{figure}[tp]
    \begin{mathparpagebreakable}
        \inferrule*
            {}{
                \boxed{- \tmcDps{\datalangRenaming} - : \datalangDef[] \rightarrow \datalangDef[] \rightarrow \Prop}
            }
        \and
        \inferrule*
            {}{
                \boxed{- \tmcDps{\datalangRenaming} - : \datalangExpr[] \times \datalangExpr[] \times \datalangExpr[] \rightarrow \datalangExpr[] \rightarrow \Prop}
            }
        \\
        \inferrule*[lab=DPSDef]
            {
                (\datalangVar_\mathit{dst}, \datalangVar_\mathit{idx}, \datalangExpr_s) \tmcDps{\datalangRenaming} \datalangExpr_t
            }{
                \datalangRec{\datalangVar}{\datalangExpr_s}
                \tmcDps{\datalangRenaming}
                \datalangRec{\datalangPair{\datalangPair{\datalangVar_\mathit{dst}}{\datalangVar_\mathit{idx}}}{\datalangVar}}{\datalangExpr_t}
            }
        \\
        \inferrule*[lab=DPSLet]
            {
                \datalangExpr_{s1} \tmcDir{\datalangRenaming} \datalangExpr_{t1}
            \and
                (\datalangExpr_\mathit{dst}, \datalangExpr_\mathit{idx}, \datalangExpr_{s2}) \tmcDps{\datalangRenaming} \datalangExpr_{t2}
            }{
                \left(
                  {\begin{array}{l}
                    \datalangExpr_\mathit{dst},
                    \datalangExpr_\mathit{idx}, \\
                    \datalangLet{\datalangVar}{\datalangExpr_{s1}}{\datalangExpr_{s2}}
                  \end{array}}
                \right)
                \tmcDps{\datalangRenaming}
                \datalangLet{\datalangVar}{\datalangExpr_{t1}}{\datalangExpr_{t2}}
            }
        \quad
        \inferrule*[lab=DPSIf]
            {
                \datalangExpr_{s0} \tmcDir{\datalangRenaming} \datalangExpr_{t0}
            \and
                (\datalangExpr_\mathit{dst}, \datalangExpr_\mathit{idx}, \datalangExpr_{s1}) \tmcDps{\datalangRenaming} \datalangExpr_{s2}
            \and
                (\datalangExpr_\mathit{dst}, \datalangExpr_\mathit{idx}, \datalangExpr_{s2}) \tmcDps{\datalangRenaming} \datalangExpr_{t2}
            }{
                \left(
                  {\begin{array}{l}
                    \datalangExpr_\mathit{dst},
                    \datalangExpr_\mathit{idx}, \\
                    \datalangIf{\datalangExpr_{s0}}{\datalangExpr_{s1}}{\datalangExpr_{s2}}
                  \end{array}}
                \right)
                \tmcDps{\datalangRenaming}
                \datalangIf{\datalangExpr_{t0}}{\datalangExpr_{t1}}{\datalangExpr_{t2}}
            }
        \\
        \inferrule*[lab=DPSBlock1]
            {
                (\datalangVar, \datalangOne, \datalangExpr_{s1}) \tmcDps{\datalangRenaming} \datalangExpr_{t1}
            \and
                \datalangExpr_{s2} \tmcDir{\datalangRenaming} \datalangExpr_{t2}
            }{
                \left(
                  {\begin{array}{l}
                    \datalangExpr_\mathit{dst},
                    \datalangExpr_\mathit{idx}, \\
                    \datalangBlock{\datalangTag}{\datalangExpr_{s1}}{\datalangExpr_{s2}}
                  \end{array}}
                \right)
                \tmcDps{\datalangRenaming}
                \begin{array}{l}
                    \datalangLet{\datalangVar}{\datalangBlock{\datalangTag}{\datalangHole}{\datalangExpr_{t2}}}{\\
                    \datalangSeq{\datalangStore{\datalangExpr_\mathit{dst}}{\datalangExpr_\mathit{idx}}{\datalangVar}}{\datalangExpr_{t1}}}
                \end{array}
            }
        \quad
        \inferrule*[lab=DPSBlock2]
            {
                \datalangExpr_{s1} \tmcDir{\datalangRenaming} \datalangExpr_{t1}
            \and
                (\datalangVar, \datalangTwo, \datalangExpr_{s2}) \tmcDps{\datalangRenaming} \datalangExpr_{t2}
            }{
                \left(
                  {\begin{array}{l}
                    \datalangExpr_\mathit{dst},
                    \datalangExpr_\mathit{idx},
                    \\
                    \datalangBlock{\datalangTag}{\datalangExpr_{s1}}{\datalangExpr_{s2}}
                  \end{array}}
                \right)
                \tmcDps{\datalangRenaming}
                \begin{array}{l}
                    \datalangLet{\datalangVar}{\datalangBlock{\datalangTag}{\datalangExpr_{t1}}{\datalangHole}}{\\
                    \datalangSeq{\datalangStore{\datalangExpr_\mathit{dst}}{\datalangExpr_\mathit{idx}}{\datalangVar}}{\datalangExpr_{t2}}}
                \end{array}
            }
        \\
        \inferrule*[lab=DPSCall]
            {
                \datalangFn \in \dom{\datalangRenaming}
            \and
                \datalangExpr_s \tmcDir{\datalangRenaming} \datalangExpr_t
            }{
                \left(
                    \datalangExpr_\mathit{dst},
                    \datalangExpr_\mathit{idx},
                    \datalangCall{\datalangFnptr{\datalangFn}}{\datalangExpr_s}
                \right)
                \tmcDps{\datalangRenaming}
                \datalangCall{
                    \datalangFnptr{\datalangRenaming [\datalangFn]}
                }{
                    \datalangPair{\datalangPair{\datalangExpr_\mathit{dst}}{\datalangExpr_\mathit{idx}}}{\datalangExpr_t}
                }
            }
        \and
        \inferrule*[lab=DPSBase]
            {
                \datalangExpr_s \tmcDir{\datalangRenaming} \datalangExpr_t
            }{
                \left(
                    \datalangExpr_\mathit{dst},
                    \datalangExpr_\mathit{idx},
                    \datalangExpr_s
                \right)
                \tmcDps{\datalangRenaming}
                \datalangStore{\datalangExpr_\mathit{dst}}{\datalangIdx}{\datalangExpr_t}
            }
    \end{mathparpagebreakable}
    \caption{Destination-passing style TMC transformation of definitions and expressions (in full)}
    \label{fig:tmc_dps}
\end{figure}


%% file: ocaml.tex
\section{\OCaml Implementation}
\label{sec:implementation}

For reasons of space, we moved some of the content in this section to \Appendices{\cref{app:more-ocaml}}{appendices}:
\Appendices{\cref{subsec:alternative-impls}}{Appendix A.1}
  discusses alternative language implementation techniques that do not require TMC.
\Appendices{\cref{subsec:design-choices}}{Appendix A.2}
  describes how the OCaml compiler decides which calls to optimize,
  and requires mandatory disambiguation hints from the user in case of ambiguity.
\Appendices{\cref{subsec:PR-history}}{Appendix A.3} provides a summary of the history of our implementation (started in 2015, restarted in 2020, merged in 2021).
\Appendices{\cref{subsec:implementation}}{Appendix A.4} explains that implementing the transformation requires a bit of care as a naive implementation is quadratic in function size. We use an applicative functor to structure a single-pass implementation that remains nicely compact and readable.
\Appendices{\cref{subsec:adoption}}{Appendix A.5} surveys the adoption of the TMC transformation in the standard library,
  and in third-party \OCaml code bases, that happened since the feature was released in 2022.

\subsection{Examples}
\label{subsec:ocaml-examples}


\begin{minipage}{0.47\linewidth}
\begin{Ocaml}
let[@tail_mod_cons] rec filter p =
  function
  | [] -> []
  | x :: xs ->
    if p x
    then x :: filter p xs
    else filter p xs
\end{Ocaml}
\end{minipage}
\hfill
\begin{minipage}{0.53\linewidth}
\begin{Ocaml}
let[@tail_mod_cons] rec merge cmp l1 l2 =
  match l1, l2 with
  | [], l | l, [] -> l
  | h1 :: t1, h2 :: t2 ->
      if cmp h1 h2 <= 0
      then h1 :: merge cmp t1 l2
      else h2 :: merge cmp l1 t2
\end{Ocaml}
\end{minipage}

TMC is not useful only for lists or other ``linear'' data types, with
at most one recursive occurrence of the datatype in each
constructor. An example follows.

\paragraph{A non-example} Consider a \ocaml{map} function on binary
trees:
\begin{Ocaml}
let[@tail_mod_cons] rec map f = function
| Leaf v -> Leaf (f v)
| Node(t1, t2) -> Node(map f t1, (map[@tailcall]) f t2)
\end{Ocaml}
In this function, there are two recursive calls, but only one of them
can be optimized; we used the \ocaml{[@tailcall]} attribute to direct
our implementation to optimize the call to the right child, as we will
discuss later. This is a \emph{bad} example of TMC usage in most
cases, given that
\begin{itemize}
\item If the tree is arbitrary, there is no reason that it would be
  right-leaning rather than left-leaning. Making only the right-child
  calls tail-calls does not protect us from stack overflows.
\item If the tree is known to be balanced, then in practice the depth
  is probably very small in both directions, so the TMC transformation
  is not necessary to have a well-behaved function.
\end{itemize}

\paragraph{Interesting non-linear examples} There \emph{are} interesting
examples of TMC-transformation on functions operating on tree-like
data structures, when there are natural assumptions about which child
is likely to contain a deep subtree. The \OCaml compiler itself
contains a number of them; consider for example the following function
from the \ocaml{Cmm} module, one of its lower-level program
representations:

\begin{Ocaml}
let[@tail_mod_cons] rec map_tail f = function
  | Clet(id, exp, body) ->
      Clet(id, exp, map_tail f body)
  | Cifthenelse(cond, ifso, ifnot) ->
      Cifthenelse(cond, map_tail f ifso, (map_tail[@tailcall]) f ifnot)
  | Csequence(e1, e2) ->
      Csequence(e1, map_tail f e2)
  | Cswitch(e, tbl, el) ->
      Cswitch(e, tbl, Array.map (map_tail f) el)
  [...]
\end{Ocaml}

This function is traversing the ``tail'' context of an arbitrary program term -- a meta-example!
The \ocaml{Cifthenelse} node acts as our binary-node constructor.
We do not know which side is likely to be larger, so TMC is not so interesting.
The recursive calls for \ocaml{Cswitch} are not in TMC position.
But on the other hand the \ocaml{Clet}, \ocaml{Csequence} do benefit from the TMC transformation: while they have several recursive subtrees, they are in practice only deeply nested in the direction that is turned into a tailcall by the transformation.
The \OCaml compiler does sometimes encounter machine-generated programs with a unusually long sequence of either constructions, and the TMC transformation may very well avoid a stack overflow in this case.

Another example would be \href{https://github.com/ocaml/ocaml/pull/9636}{\#9636}, a patch to the \OCaml compiler proposed in June 2020 by Mark Shinwell, to get a partially-tail-recursive implementation of the ``Common Subexpression Elimination'' (CSE) pass through a manual continuation-passing-style transform.
Xavier Leroy remarked that the existing implementation in fact fits the TMC fragment. Not all recursive calls become tail-calls (this would require a more powerful transformation or a longer, less readable patch), but the behavior of TMC on the unchanged code matches the tail-call-ness proposed in the human-written patch.

\subsection{Specifying Which Calls are in TMC Position} \label{subsec:specification}
To reason about the stack usage of their programs, users must understand which calls are in tail-modulo-cons position.
Informally, they are the calls placed under any composition of either tail-recursive or constructor contexts.

We can in fact give a simple formal description of this intuition, here for \DataLang in \cref{fig:contexts}.
A tail frame $\datalangTailFrame$ is a single term-former with holes in tail-position.
A constructor frame $\datalangConsFrame$ is a single constructor term-former (we omit deterministic blocks, which do not occur in the source).
A tail context $\datalangTailCtx$ is an arbitrary composition of tail-frame, and a TMC context $\datalangTMCCtx$ is an arbitrary composition of tail frames and constructor frames.

If a source function can be decomposed in a TMC context $\datalangTMCCtx$ with source expressions in its holes, some of which are calls to TMC-transformed functions, then our relation admits a DPS transformation where all those function calls are tail-calls, and this transformation is reachable in our \OCaml implementation, possibly by adding some annotations.

Note in particular that we do not only optimize calls to the same function we are defining, direct calls to arbitrary other functions can be transformed, if those functions have been annotated to be TMC-transformed.
This is analogous to how most functional languages support arbitrary \emph{tail calls} and not just tail self-recursion. We seamlessly support mutually recursive functions, DPS calls into locally-bound functions, etc. On the other hand, we currently do not optimize call to higher-order function arguments, or calls crossing module boundaries.

\begin{figure}[tp]
    \begin{tabular}{lclcl}
            $\datalangTailFrame[]$
            & $\ni$ &
            $\datalangTailFrame$
            & $\Coloneqq$ &
            $\datalangLet {\datalangVar} {\datalangExpr} \datalangCtxHole \mid \datalangIf \datalangExpr \datalangCtxHole \datalangCtxHole$
\\
            $\datalangConsFrame[]$
            & $\ni$ &
            $\datalangConsFrame$
            & $\Coloneqq$ &
            $\datalangBlock \datalangTag \datalangExpr \datalangCtxHole
            \mid \datalangBlock \datalangTag \datalangCtxHole \datalangExpr$
\\
            $\datalangTMCFrame[]$
            & $\ni$ &
            $\datalangTMCFrame$
            & $\Coloneqq$ &
            $\datalangTailFrame
             \mid
             \datalangConsFrame$
\\
            $\datalangTailCtx[]$
            & $\ni$ &
            $\datalangTailCtx$
            & $\Coloneqq$ &
            $\datalangCtxHole
             \mid
             \datalangTailFrame
             \mid
             \datalangTailCtx{[\datalangTailCtx]}$
\\
            $\datalangTMCCtx[]$
            & $\ni$ &
            $\datalangTMCCtx$
            & $\Coloneqq$ &
            $\datalangCtxHole
             \mid
             \datalangTMCFrame
             \mid
             \datalangTMCCtx{[\datalangTMCCtx]}$
    \end{tabular}
    \caption{\DataLang contexts for optimizable calls}
    \label{fig:contexts}
\end{figure}

\subsection{Constructor Compression} \label{subsec:constructor-compression} The translation as we described it formally in \cref{subsec:transformation} generates unpleasant code when many constructors are nested before the recursive call. For example, consider this strange function duplicating each element of a list:
\begin{Ocaml}
let rec dup = function [] -> [] | x :: xs -> x :: x :: dup xs
\end{Ocaml}

Such nested constructors are common in compiler code bases, for
example a desugaring pass that transforms a single term-former into
a composition of several simpler term-formers, and applies recursively to
its subterms.

Following the TMC transformation naively, the DPS version would propagate two different locations and performs two writes. We introduced ``constructor compression'', an optimization of the generated code that avoids creating intermediary destinations for nested constructors, leading to clearer generated code and better constant factors. Compare the naive translation of \ocaml|dup|, on the left, and our compressed translation on the right:

\begin{minipage}{0.5\linewidth}
\begin{Ocaml}
let rec dup_dps dst ofs = function
| [] -> dst.(ofs) <- []
| x :: xs ->
  let dst1 = x :: ? in
  dst.(ofs) <- dst1;
  let dst2 = x :: ? in
  dst1.(1) <- dst2;
  dup_dps dst2 1 xs
\end{Ocaml}
\end{minipage}
\hfill
\begin{minipage}{0.5\linewidth}
\begin{Ocaml}
let rec dup_dps dst ofs = function
| [] -> dst.(ofs) <- []
| x :: xs ->
  let dst2 = x :: ? in
  dst.(ofs) <- x :: dst2;
  dup_dps dst2 1 xs
\end{Ocaml}
\end{minipage}

This is implemented by passing a new transformation parameter: a stack of ``delayed'' constructor applications, that are in context and must be applied to the result of the subterm. When we encounter the final recursive call, we ``reify'' this stack: the last/innermost constructor in the stack becomes the new destination (\ocaml|dst2| in the example above), and the rest of the stack is applied to the new destination when we write to the old destination. There are two subtleties:
\begin{enumerate}
\item \ocaml|if p then e1 else e2| has two subterms which are transformed in DPS style, and naively passing the stack of delayed constructors to both subterms would duplicate code; instead we also reify the current stack when encountering such constructs. For example,\\
\ocaml{  1 :: 2 :: if p then (3 :: f ()) else (4 :: f ())}%
becomes:\\
\ocaml{  let dst1 = 1 :: 2 :: ? in}\\
\ocaml{  if p then (let dst2 = 3 :: ? in dst1.1 <- dst2; f dst2 1 ())}\\
\ocaml{       else (let dst3 = 4 :: ? in dst1.1 <- dst3; f dst3 1 ())}

\item This transformation may permute constructor applications after effectful subterms. If the constructor application context frame contains possibly-effectful subterms (for example \ocaml{f x :: _? } instead of \ocaml{x :: _?}), the compiler must \ocaml|let|-bind them at their original position to avoid changing the evaluation order. For example,\\
\ocaml{x () :: (y (); f ())} does not become \ocaml{y (); (let dst = x () :: ? in ...)},
but instead \ocaml{let tmp = x () in (y (); let dst = tmp :: ? in ...)}.
\end{enumerate}

It is conceptually easy to extend our previous formalization of TMC as a rewriting relation to capture constructor compression, by indexing this relation on an additional list of constructor contexts. We do not present this here for lack of space, but included this change in our \Coq proofs, which establish correctness of TMC in presence of constructor compression.

\subsection{Evaluation: Benchmarks}

We measured the performance of \ocaml|List.map (fun n -> n + 1)| to validate our claims that the TMC transformation preserves program performance, and lets us replace complex hand-optimized tail-recursive implementations. \ocaml|List.map| is a worst-case: with most of the time spent in recursion and list construction, it is more sensitive to constant-factor overheads than other recursive functions.

\begin{figure}[tp]
\def\svgscale{0.8}
\graphicspath{{plots/}}
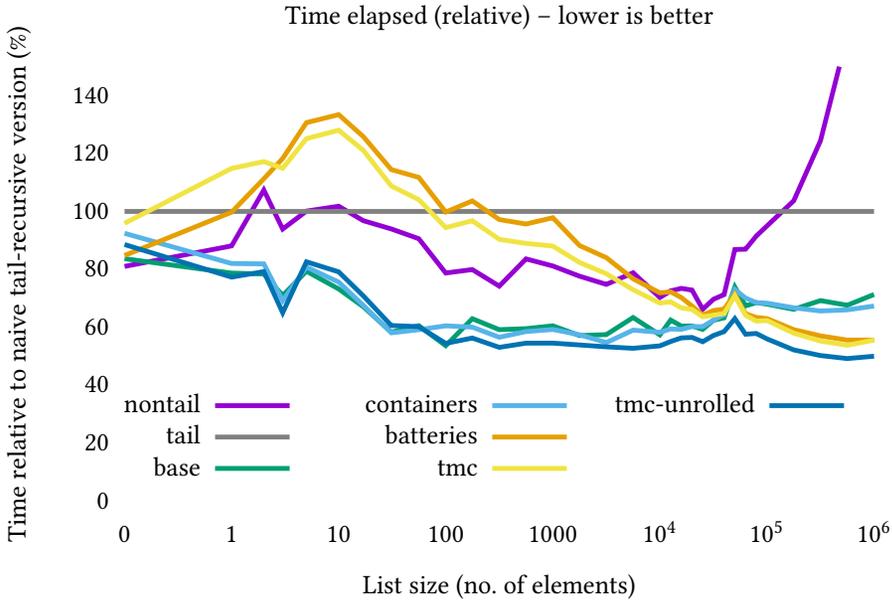
\caption{\ocaml|List.map| benchmark on \OCaml~5.1}
\label{fig:bench5}
\end{figure}

\newcommand{\bench}[1]{\textbf{#1}}

The different versions we benchmark are the following. We measure the code size (in lines) of each version, as a reasonable approximation of its implementation complexity.
\begin{description}
\item[\bench{nontail}] (5 lines of code) The naive, non-tail-recursive implementation.
\item[\bench{tail}] (9 lines) The naive tail-recursive implementation,
  \ocaml{List.rev (List.rev_map f xs)}.
\item[\bench{base}] (78 lines) The implementation of Jane Street's
  \href{https://github.com/janestreet/base}{Base} library
  (version 0.14.0). It is heavily hand-optimized to compensate for the costs
  of being tail-recursive.
\item[\bench{containers}] (55 lines) Another standard-library extension by Simon
  Cruanes; it is the hand-optimized tail-recursive implementation we
  included in the Prologue.
\item[\bench{batteries}] (29 lines) The implementation of the community-maintained
  \href{https://github.com/ocaml-batteries-team/batteries-included/}{Batteries}
  library. It is actually written in destination-passing-style, using
  an unsafe encoding with \ocaml{Obj.magic} to unsafely cast a mutable
  record into a list cell. (The trick comes from the older Extlib
  library, was introduced by Brian Hurt in 2003, and has a comment crediting Jacques
  Garrigue for the particular encoding used.)
\item[\bench{tmc}] (5 lines) ``Our'' version, the last version of the Prologue: the
  result of applying our implementation of the TMC transformation to
  the simple, non-tail-recursive version.
\item[\bench{tmc-unrolled}] (18 lines) The result of manually unrolling the \bench{tmc} implementation three times, to be compared with \bench{base} and \bench{containers} that use manual unrolling as well.
\end{description}

The benchmarks reports the relative performance compared to the naive tail-recursive version as our baseline. They were run on \OCaml~5.1 in July 2024, on a Linux machine with an AMD Ryzen processor fixed at a 3Ghz frequency, looping each measurement for 5s (a single \ocaml|List.map| run takes between 7ns, for empty lists, and 89ms on lists with a million element).

Qualitatively we see that there are four groups:
\begin{itemize}
\item \bench{tmc}, \bench{batteries} perform very well on large lists, but they
  are slower than the baseline on small lists.
\item \bench{nontail} performs better than \bench{tmc}, \bench{batteries} on list sizes up to $10^4$, and much worse on larger lists.
\item \bench{base}, \bench{containers} perform noticeably better than \bench{nontail} at all sizes,
  but worse than the TMC versions above size $10^4$.
\item \bench{tmc-unrolled} is the best option: it performs as well as \bench{base} and \bench{containers} before $10^4$, and as well as \bench{tmc}, \bench{batteries} afterwards.
\end{itemize}
Our interpretation of the result is that some unrolling makes a noticeable performance difference for such a short function: \bench{tmc} is not good enough on smaller lists, but \bench{tmc-unrolled} is the best-performing, despite being much simpler than the \bench{base} and \bench{containers} versions.

\paragraph{Asymptotics of \bench{nontail}} The bad behavior of \bench{nontail} on large lists comes from a quadratic behavior on very large call stacks, coming from a repeated scan of the call stack during minor collections. (The \OCaml compiler and runtime could be tweaked to avoid this quadratic behavior, at the cost of some small constant overhead on function returns.)



%% file: plots/plot.5.pdf_tex
\begingroup%
  \makeatletter%
  \providecommand\color[2][]{%
    \errmessage{(Inkscape) Color is used for the text in Inkscape, but the package 'color.sty' is not loaded}%
    \renewcommand\color[2][]{}%
  }%
  \providecommand\transparent[1]{%
    \errmessage{(Inkscape) Transparency is used (non-zero) for the text in Inkscape, but the package 'transparent.sty' is not loaded}%
    \renewcommand\transparent[1]{}%
  }%
  \providecommand\rotatebox[2]{#2}%
  \newcommand*\fsize{\dimexpr\f@size pt\relax}%
  \newcommand*\lineheight[1]{\fontsize{\fsize}{#1\fsize}\selectfont}%
  \ifx\svgwidth\undefined%
    \setlength{\unitlength}{450bp}%
    \ifx\svgscale\undefined%
      \relax%
    \else%
      \setlength{\unitlength}{\unitlength * \real{\svgscale}}%
    \fi%
  \else%
    \setlength{\unitlength}{\svgwidth}%
  \fi%
  \global\let\svgwidth\undefined%
  \global\let\svgscale\undefined%
  \makeatother%
  \begin{picture}(1,0.66666667)%
    \lineheight{1}%
    \setlength\tabcolsep{0pt}%
    \put(0,0){\includegraphics[width=\unitlength,page=1]{plot.5.pdf}}%
    \put(0.12473333,0.10441667){\makebox(0,0)[rt]{\lineheight{1.25}\smash{\begin{tabular}[t]{r}0\end{tabular}}}}%
    \put(0,0){\includegraphics[width=\unitlength,page=2]{plot.5.pdf}}%
    \put(0.12473333,0.16436667){\makebox(0,0)[rt]{\lineheight{1.25}\smash{\begin{tabular}[t]{r}20\end{tabular}}}}%
    \put(0,0){\includegraphics[width=\unitlength,page=3]{plot.5.pdf}}%
    \put(0.12473333,0.22431667){\makebox(0,0)[rt]{\lineheight{1.25}\smash{\begin{tabular}[t]{r}40\end{tabular}}}}%
    \put(0,0){\includegraphics[width=\unitlength,page=4]{plot.5.pdf}}%
    \put(0.12473333,0.28428333){\makebox(0,0)[rt]{\lineheight{1.25}\smash{\begin{tabular}[t]{r}60\end{tabular}}}}%
    \put(0,0){\includegraphics[width=\unitlength,page=5]{plot.5.pdf}}%
    \put(0.12473333,0.34423333){\makebox(0,0)[rt]{\lineheight{1.25}\smash{\begin{tabular}[t]{r}80\end{tabular}}}}%
    \put(0,0){\includegraphics[width=\unitlength,page=6]{plot.5.pdf}}%
    \put(0.12473333,0.40418333){\makebox(0,0)[rt]{\lineheight{1.25}\smash{\begin{tabular}[t]{r}100\end{tabular}}}}%
    \put(0,0){\includegraphics[width=\unitlength,page=7]{plot.5.pdf}}%
    \put(0.12473333,0.46413333){\makebox(0,0)[rt]{\lineheight{1.25}\smash{\begin{tabular}[t]{r}120\end{tabular}}}}%
    \put(0,0){\includegraphics[width=\unitlength,page=8]{plot.5.pdf}}%
    \put(0.12473333,0.52408333){\makebox(0,0)[rt]{\lineheight{1.25}\smash{\begin{tabular}[t]{r}140\end{tabular}}}}%
    \put(0,0){\includegraphics[width=\unitlength,page=9]{plot.5.pdf}}%
    \put(0.14105,0.06941667){\makebox(0,0)[t]{\lineheight{1.25}\smash{\begin{tabular}[t]{c}0\end{tabular}}}}%
    \put(0,0){\includegraphics[width=\unitlength,page=10]{plot.5.pdf}}%
    \put(0.25326667,0.06941667){\makebox(0,0)[t]{\lineheight{1.25}\smash{\begin{tabular}[t]{c}1\end{tabular}}}}%
    \put(0,0){\includegraphics[width=\unitlength,page=11]{plot.5.pdf}}%
    \put(0.36548333,0.06941667){\makebox(0,0)[t]{\lineheight{1.25}\smash{\begin{tabular}[t]{c}10\end{tabular}}}}%
    \put(0,0){\includegraphics[width=\unitlength,page=12]{plot.5.pdf}}%
    \put(0.4777,0.06941667){\makebox(0,0)[t]{\lineheight{1.25}\smash{\begin{tabular}[t]{c}100\end{tabular}}}}%
    \put(0,0){\includegraphics[width=\unitlength,page=13]{plot.5.pdf}}%
    \put(0.58993333,0.06941667){\makebox(0,0)[t]{\lineheight{1.25}\smash{\begin{tabular}[t]{c}1000\end{tabular}}}}%
    \put(0,0){\includegraphics[width=\unitlength,page=14]{plot.5.pdf}}%
    \put(0.70215,0.06941667){\makebox(0,0)[t]{\lineheight{1.25}\smash{\begin{tabular}[t]{c}$10^4$\end{tabular}}}}%
    \put(0,0){\includegraphics[width=\unitlength,page=15]{plot.5.pdf}}%
    \put(0.81436667,0.06941667){\makebox(0,0)[t]{\lineheight{1.25}\smash{\begin{tabular}[t]{c}$10^5$\end{tabular}}}}%
    \put(0,0){\includegraphics[width=\unitlength,page=16]{plot.5.pdf}}%
    \put(0.92658333,0.06941667){\makebox(0,0)[t]{\lineheight{1.25}\smash{\begin{tabular}[t]{c}$10^6$\end{tabular}}}}%
    \put(0.0373,0.33681666){\rotatebox{90}{\makebox(0,0)[t]{\lineheight{1.25}\smash{\begin{tabular}[t]{c}Time relative to naive tail-recursive version (\%)\end{tabular}}}}}%
    \put(0.53381666,0.01691667){\makebox(0,0)[t]{\lineheight{1.25}\smash{\begin{tabular}[t]{c}List size (no. of elements)\end{tabular}}}}%
    \put(0.22076667,0.2037){\makebox(0,0)[rt]{\lineheight{1.25}\smash{\begin{tabular}[t]{r}nontail\end{tabular}}}}%
    \put(0,0){\includegraphics[width=\unitlength,page=17]{plot.5.pdf}}%
    \put(0.22076667,0.1712){\makebox(0,0)[rt]{\lineheight{1.25}\smash{\begin{tabular}[t]{r}tail\end{tabular}}}}%
    \put(0,0){\includegraphics[width=\unitlength,page=18]{plot.5.pdf}}%
    \put(0.22076667,0.1387){\makebox(0,0)[rt]{\lineheight{1.25}\smash{\begin{tabular}[t]{r}base\end{tabular}}}}%
    \put(0,0){\includegraphics[width=\unitlength,page=19]{plot.5.pdf}}%
    \put(0.51126667,0.2037){\makebox(0,0)[rt]{\lineheight{1.25}\smash{\begin{tabular}[t]{r}containers\end{tabular}}}}%
    \put(0,0){\includegraphics[width=\unitlength,page=20]{plot.5.pdf}}%
    \put(0.51126667,0.1712){\makebox(0,0)[rt]{\lineheight{1.25}\smash{\begin{tabular}[t]{r}batteries\end{tabular}}}}%
    \put(0,0){\includegraphics[width=\unitlength,page=21]{plot.5.pdf}}%
    \put(0.51126667,0.1387){\makebox(0,0)[rt]{\lineheight{1.25}\smash{\begin{tabular}[t]{r}tmc\end{tabular}}}}%
    \put(0,0){\includegraphics[width=\unitlength,page=22]{plot.5.pdf}}%
    \put(0.80176667,0.2037){\makebox(0,0)[rt]{\lineheight{1.25}\smash{\begin{tabular}[t]{r}tmc-unrolled\end{tabular}}}}%
    \put(0,0){\includegraphics[width=\unitlength,page=23]{plot.5.pdf}}%
    \put(0.53381667,0.60656667){\makebox(0,0)[t]{\lineheight{1.25}\smash{\begin{tabular}[t]{c}Time elapsed (relative) – lower is better\end{tabular}}}}%
  \end{picture}%
\endgroup%

%% file: specification.tex
\section{Specifying TMC}
\label{sec:specification}

In this section, we gradually introduce aspects of our relational separation logic, by introducing our specifications for the direct-style and destination-passing-style transformations of \cref{sec:formalization} in relational separation logic.

\subsection{Direct Transformation}

Intuitively, the direct transformation $\datalangExpr_s \tmcDir{\datalangRenaming} \datalangExpr_t$ preserves the behaviors of the source expression $\datalangExpr_s$.
Basically, $\datalangExpr_s$ and $\datalangExpr_t$ compute the same thing.
Using \emph{relational Hoare logic}\Xfrancois{TODO cite Benton 2004}, a extension of standard Hoare logic relating two expressions, we would write:
\[
    \iSimvHoare{
        \datalangExpr_s \tmcDir{\datalangRenaming} \datalangExpr_t
    }{
        \datalangVal_s, \datalangVal_t \ldotp
        \datalangVal_s \iSimilar \datalangVal_t
    }{
        \datalangExpr_s
    }{
        \datalangExpr_t
    }
\]
\Xfrancois{The $\datalangExpr_s \tmcDir{\datalangRenaming} \datalangExpr_t$ is pure and could be outside the triple.}

The informal meaning of this specification is that 1) $\datalangExpr_t$ refines $\datalangExpr_s$ in the sense that any behavior (converging, diverging or stuck execution) of $\datalangExpr_t$ is also a behavior of $\datalangExpr_s$ and 2) if $\datalangExpr_t$ converges to value $\datalangVal_t$, then $\datalangExpr_s$ also converges to some value $\datalangVal_s$ that is \emph{similar} to $\datalangVal_s$.
We will formalize the notion of \emph{behavior} in \cref{sec:simulation} and that of \emph{similarity} later in this section.
For the time being, the reader may assume similarity is just equality on values.

\subsection{DPS Transformation}

The DPS transformation $(\datalangLoc, \datalangIdx, \datalangExpr_s) \tmcDps{\datalangRenaming} \datalangExpr_t$ is parameterized by a destination $(\datalangLoc, \datalangIdx)$ pointing to an uninitialized field of some block.
Intuitively, $\datalangExpr_t$ computes the same thing as $\datalangExpr_s$ but writes it into the destination instead of returning it. This can be expressed concisely in \emph{relational separation logic}~\citep*{yang-07}, a further extension of relational Hoare logic:
\[
    \iSimvHoare{
        (\datalangLoc, \datalangIdx, \datalangExpr_s) \tmcDps{\datalangRenaming} \datalangExpr_t \iSep
        (\datalangLoc + \datalangIdx) \iPointsto_t \datalangHole
    }{
        \datalangVal_s, \datalangUnit \ldotp
        \exists \datalangVal_t \ldotp
        (\datalangLoc + \datalangIdx) \iPointsto_t \datalangVal_t \iSep
        \datalangVal_s \iSimilar \datalangVal_t
    }{
        \datalangExpr_s
    }{
        \datalangExpr_t
    }
\]

In words: if $\datalangExpr_s$ transforms into $\datalangExpr_t$, and if we uniquely own the destination location $\datalangLoc + \datalangIdx$, we can transfer ownership to $\datalangExpr_t$ and run the two programs, whose execution must be related. When they reduce to values, $\datalangExpr_s$ reduces to a source value $\datalangVal_s$ and $\datalangExpr_t$ to the unit value $\datalangUnit$, and we recover the unique ownership of the destination, which now contains a target value $\datalangVal_t$ similar to $\datalangVal_s$.

\subsection{Heap Bijection}

Defining value similarity as just syntactic equality is not sufficient: corresponding source and target block allocations are not done in lockstep, so the resulting locations may differ.
For example, consider the \datalang{map} function and its DPS transform from \cref{subsec:tmc_example}.
In the source program, the cons cell \datalang{y :: @map (fn, xs)} is allocated after the recursive call.
In the transformed program, the corresponding block is allocated before the call.

To deal with this, we introduce a \emph{heap bijection} as in \Simuliris~\citep*{simuliris-2022}.
This is a partial bijection (some destination locations have no source counterpart) which grows over time.
Its usage is formalized by the \RefTirName{BijInsert} rule:

\begin{mathline}
    \inferrule*[lab=BijInsert]
        {
            \datalangLoc_s \iPointsto_s \datalangVal_s
        \and
            \datalangLoc_t \iPointsto_t \datalangVal_t
        \and
            \datalangVal_s \iSimilar \datalangVal_t
        }{
            \datalangLoc_s \iInBij \datalangLoc_t
        }
\end{mathline}
This is a \emph{ghost update} rule that mutates the logical state.
It can only be applied when the two locations $\datalangLoc_s$ and $\datalangLoc_t$ have similar content $\datalangVal_s \iSimilar \datalangVal_t$.
It consumes the ``private'' ownership of the source and target points-to $\datalangLoc_s \iPointsto_s \datalangVal_s$ and $\datalangLoc_t \iPointsto_t \datalangVal_t$, and produces a persistent proposition $\datalangLoc_s \iInBij \datalangLoc_t$ witnessing that the two locations are now in the ``public'' bijection.

We formally define value similarity $\datalangVal_s \iSimilar \datalangVal_t$ in \cref{fig:isimilar}.
It coincides with equality except on blocks, for which we require all fields to be registered in the bijection.

\input{figures/isimilar}


%% file: figures/isimilar.tex
\begin{figure}[tp]
    \centering
    \begin{mathparpagebreakable}
        \inferrule*
            {}{
                \datalangUnit \iSimilar \datalangUnit
            }
        \and
        \inferrule*
            {}{
                \datalangIdx \iSimilar \datalangIdx
            }
        \and
        \inferrule*
            {}{
                \datalangTag \iSimilar \datalangTag
            }
        \and
        \inferrule*
            {}{
                \datalangBool \iSimilar \datalangBool
            }
        \and
        \inferrule*
            {
                \forall \datalangIdx \in \datalangIdx[] \ldotp
                (\datalangLoc_s + \datalangIdx) \iInBij (\datalangLoc_t + \datalangIdx)
            }{
                \datalangLoc_s \iSimilar \datalangLoc_t
            }
        \and
        \inferrule*
            {
                \datalangFn \in \dom{\datalangProg_s}
            }{
                \datalangFnptr{\datalangFn} \iSimilar \datalangFnptr{\datalangFn}
            }
    \end{mathparpagebreakable}
    \caption{Similarity in $\iProp$}
    \label{fig:isimilar}
\end{figure}

%% file: program_logic.tex
\section{Relational separation logic}
\label{sec:program_logic}

\input{figures/program_logic}

In this section, we describe our relational program logic, presented in \cref{fig:program_logic}.
We omit some congruence rules for brevity.
The relation $\iSimv[\iProt]{\iPred}{\datalangExpr_s}{\datalangExpr_t}$ relates a source expression $\datalangExpr_s$ with a target expression $\datalangExpr_t$ under a postcondition $\iPred$, following the protocol $\iProt$. Informally, this is a backward simulation: any execution of the target term $\datalangExpr_t$ can be mapped back to an execution of the source term $\datalangExpr_s$, and if the target term reaches a value $\datalangVal_t$ then the source term can reach a $\datalangVal_s$ such that the postcondition $\Phi(\datalangVal_s, \datalangVal_t)$ holds. The protocol $\iProt$ specifies pairs of abstract transitions, that could model foreign/external calls for example, that have to be taken in lockstep on both side. Formally, the judgment $\iSimv[\iProt]{\iPred}{\datalangExpr_s}{\datalangExpr_t}$ in our program logic establishes a simulation relation $\mathrm{\color{\iSimGfpColor}{sim}}_\iProt (\iPred, \datalangExpr_s, \datalangExpr_t)$ that we will define in \cref{sec:simulation}.\footnote{As usual, the relation between the program logic and the simulation can be viewed in two ways. You can view the program logic as a syntactic system of inference rules, with a proof that if a judgment admits a closed derivation then the corresponding simulation statement holds. Or you can think of the program logic judgment and the simulation statement as the same thing, and inference rules are a convenient notation for admissibility lemmas.}

We extend it to support a precondition in the standard way:
\begin{mathline}
    \iSimvHoare[\iProt]{P}{\iPred}{\datalangExpr_s}{\datalangExpr_t}
    \coloneqq
    \iPersistent \left( P \iWand \iSimv[\iProt]{\iPred}{\datalangExpr_s}{\datalangExpr_t} \right)
\end{mathline}
Compared to the specifications of \cref{sec:specification}, we introduced an additional protocol parameter $\iProt$.
We explain it together with the \RefTirName{RelProtocol} rule in \cref{subsec:protocols}.

\paragraph{Language-independent rules}
The following rules are independent of \DataLang and could be reused as is in further works.

\RefTirName{RelPost} states that two values are related when they are in the relational postcondition.

\RefTirName{RelStuck} relates \emph{strongly stuck} expressions.
\Xfrancois{Explain up front that your wp is non-standard as it does not guarantee safety.}
An expression is strongly stuck when it is stuck for any heap state.
\Xfrancois{Explain why ``strongly'' is needed here.}

\RefTirName{RelBind} is a standard bind rule sequencing computations on both sides.

\RefTirName{RelSrcPure} and \RefTirName{RelTgtPure} let us take pure reduction steps in either the source or target.
Pure steps (definition omitted for brevity) are the reduction steps that are deterministic and do not depend on the state.

\paragraph{Language-specific rules: non-determinism}
$\iSimv[\iProt]{\iPred}{\datalangExpr_s}{\datalangExpr_t}$ asserts that $\datalangExpr_t$ refines $\datalangExpr_s$: any behavior of $\datalangExpr_t$ is also a behavior of $\datalangExpr_s$.
Consequently, non-determinism is treated differently in the source and target: we treat non-determinism as \emph{angelic} in source reductions and \emph{demonic} in target reductions.

Our operational semantics uses non-determinism in the reduction of constructors: $\datalangBlock{\datalangTag}{\datalangExpr_1}{\datalangExpr_2}$ reduces to $\datalangBlockDet{\datalangTag}{\datalangVar_1}{\datalangVar_2}$, where $\datalangVar_1$ and $\datalangVar_2$ are bound to $\datalangExpr_1$ and $\datalangExpr_2$ in some non-deterministic order.
In the program logic, the user may \emph{choose} an order for the source reduction, by using one of the rules \RefTirName{RelSrcBlock1} or \RefTirName{RelSrcBlock2}. On the other hand, they have to prove that the expressions are related against \emph{any} target order, by proving the two premises of the rule \RefTirName{RelTgtBlock}.

\paragraph{Language-specific rules: private locations.}
We can reason on points-to assertions in a standard way.
From a deterministic constructor $\datalangBlockDet{\datalangTag}{\datalangVal_1}{\datalangVal_2}$, we can apply \RefTirName{RelSrcBlockDet} or \RefTirName{RelTgtBlockDet}, yielding a points-to assertion for the allocated block.
The rules \RefTirName{RelSrcLoad} and \RefTirName{RelTgtLoad} let us load the pointed value while \RefTirName{RelSrcStore} and \RefTirName{RelTgtStore} let us update it with a new value.
We interpret locations owned by a points-to assertion as ``private'' to the source or target: they are not registered in the ``public'' partial heap bijection.

\paragraph{Language-specific rules: locations in the bijection.}
Corresponding source and target locations registered in the bijection through \RefTirName{BijInsert} have given up their respective points-to assertions but can still be accessed using the rules \RefTirName{RelLoad} and \RefTirName{RelStore}.

\RefTirName{RelLoad} states that simultaneously loading from two corresponding blocks yields similar values.

\RefTirName{RelStore} lets us store similar values into the same field of two corresponding blocks.

These two rules enforce the bijection invariant: corresponding blocks contain similar values.


%% file: figures/program_logic.tex
\begin{figure}[tp]
    \begin{mathparpagebreakable}
        \inferrule*[lab=RelPost]
            {
                \iPred (\datalangVal_s, \datalangVal_t)
            }{
                \iSimv[\iProt]{\iPred}{\datalangVal_s}{\datalangVal_t}
            }
        \and
        \inferrule*[lab=RelStuck]
            {
                \mathrm{strongly \mathhyphen stuck}_{\datalangProg_s} (\datalangExpr_s)
            \and
                \mathrm{strongly \mathhyphen stuck}_{\datalangProg_t} (\datalangExpr_t)
            }{
                \iSimv[\iProt]{\iPred}{\datalangExpr_s}{\datalangExpr_t}
            }
        \\
        \inferrule*[lab=RelBind]
            {
                \iSimv[\iProt]{\lambdaAbs (\datalangVal_s, \datalangVal_t) \ldotp \iSimv[\iProt]{\iPred}{\datalangEctx_s [\datalangVal_s]}{\datalangEctx_t [\datalangVal_t]}}{\datalangExpr_s}{\datalangExpr_t}
            }{
                \iSimv[\iProt]{\iPred}{\datalangEctx_s [\datalangExpr_s]}{\datalangEctx_t [\datalangExpr_t]}
            }
        \\
        \inferrule*[lab=RelSrcPure]
            {
                \datalangExpr_s \pureStep{\datalangProg_s} \datalangExpr_s'
            \and
                \iSimv[\iProt]{\iPred}{\datalangExpr_s'}{\datalangExpr_t}
            }{
                \iSimv[\iProt]{\iPred}{\datalangExpr_s}{\datalangExpr_t}
            }
        \and
        \inferrule*[lab=RelTgtPure]
            {
                \datalangExpr_t \pureStep{\datalangProg_t} \datalangExpr_t'
            \and
                \iSimv[\iProt]{\iPred}{\datalangExpr_s}{\datalangExpr_t'}
            }{
                \iSimv[\iProt]{\iPred}{\datalangExpr_s}{\datalangExpr_t}
            }
        \\
        \inferrule*[lab=RelSrcBlock1]
            {
                \iSimv[\iProt]{\iPred}{
                    \begin{array}{l}
                        \datalangLet{\datalangVar_1}{\datalangExpr_{s1}}{\\
                        \datalangLet{\datalangVar_2}{\datalangExpr_{s2}}{\\
                        \datalangBlockDet{\datalangTag}{\datalangVar_1}{\datalangVar_2}}}
                    \end{array}
                }{\datalangExpr_t}
            }{
                \iSimv[\iProt]{\iPred}{\datalangBlock{\datalangTag}{\datalangExpr_{s1}}{\datalangExpr_{s2}}}{\datalangExpr_t}
            }
        \and
        \inferrule*[lab=RelSrcBlock2]
            {
                \iSimv[\iProt]{\iPred}{
                    \begin{array}{l}
                        \datalangLet{\datalangVar_2}{\datalangExpr_{s2}}{\\
                        \datalangLet{\datalangVar_1}{\datalangExpr_{s1}}{\\
                        \datalangBlockDet{\datalangTag}{\datalangVar_1}{\datalangVar_2}}}
                    \end{array}
                }{\datalangExpr_t}
            }{
                \iSimv[\iProt]{\iPred}{\datalangBlock{\datalangTag}{\datalangExpr_{s1}}{\datalangExpr_{s2}}}{\datalangExpr_t}
            }
        \and
        \inferrule*[lab=RelTgtBlock]
            {
                \iSimv[\iProt]{\iPred}{\datalangExpr_s}{
                    \begin{array}{l}
                        \datalangLet{\datalangVar_1}{\datalangExpr_{t1}}{\\
                        \datalangLet{\datalangVar_2}{\datalangExpr_{t2}}{\\
                        \datalangBlockDet{\datalangTag}{\datalangVar_1}{\datalangVar_2}}}
                    \end{array}
                }
            \and
                \iSimv[\iProt]{\iPred}{\datalangExpr_s}{
                    \begin{array}{l}
                        \datalangLet{\datalangVar_2}{\datalangExpr_{t2}}{\\
                        \datalangLet{\datalangVar_1}{\datalangExpr_{t1}}{\\
                        \datalangBlockDet{\datalangTag}{\datalangVar_1}{\datalangVar_2}}}
                    \end{array}
                }
            }{
                \iSimv[\iProt]{\iPred}{\datalangExpr_s}{\datalangBlock{\datalangTag}{\datalangExpr_{t1}}{\datalangExpr_{t2}}}
            }
        \\
        \inferrule*[lab=RelSrcBlockDet]
            {
                \forall \datalangLoc_s \ldotp
                \datalangLoc_s \iPointsto_s (\datalangTag, \datalangVal_{s1}, \datalangVal_{s2}) \iWand
                \iSimv[\iProt]{\iPred}{\datalangLoc_s}{\datalangExpr_t}
            }{
                \iSimv[\iProt]{\iPred}{\datalangBlockDet{\datalangTag}{\datalangVal_{s1}}{\datalangVal_{s2}}}{\datalangExpr_t}
            }
        \and
        \inferrule*[lab=RelTgtBlockDet]
            {
                \forall \datalangLoc_t \ldotp
                \datalangLoc_t \iPointsto_t (\datalangTag, \datalangVal_{t1}, \datalangVal_{t2}) \iWand
                \iSimv[\iProt]{\iPred}{\datalangExpr_s}{\datalangLoc_t}
            }{
                \iSimv[\iProt]{\iPred}{\datalangExpr_s}{\datalangBlockDet{\datalangTag}{\datalangVal_{t1}}{\datalangVal_{t2}}}
            }
        \\
        \inferrule*[lab=RelSrcLoad]
            {
                (\datalangLoc_s + \datalangIdx) \iPointsto_s \datalangVal_s
            \\\\
                (\datalangLoc_s + \datalangIdx) \iPointsto_s \datalangVal_s \iWand
                \iSimv[\iProt]{\iPred}{\datalangVal_s}{\datalangExpr_t}
            }{
                \iSimv[\iProt]{\iPred}{\datalangLoad{\datalangLoc_s}{\datalangIdx}}{\datalangExpr_t}
            }
        \and
        \inferrule*[lab=RelTgtLoad]
            {
                (\datalangLoc_t + \datalangIdx) \iPointsto_t \datalangVal_t
            \\\\
                (\datalangLoc_s + \datalangIdx) \iPointsto_t \datalangVal_t \iWand
                \iSimv[\iProt]{\iPred}{\datalangExpr_s}{\datalangVal_t}
            }{
                \iSimv[\iProt]{\iPred}{\datalangExpr_s}{\datalangLoad{\datalangLoc_t}{\datalangIdx}}
            }
        \\
        \inferrule*[lab=RelSrcStore]
            {
                (\datalangLoc_s + \datalangIdx) \iPointsto_s \datalangVal_s
            \\\\
                (\datalangLoc_s + \datalangIdx) \iPointsto_s \datalangVal_s' \iWand
                \iSimv[\iProt]{\iPred}{\datalangUnit}{\datalangExpr_t}
            }{
                \iSimv[\iProt]{\iPred}{\datalangStore{\datalangLoc_s}{\datalangIdx}{\datalangVal_s'}}{\datalangExpr_t}
            }
        \and
        \inferrule*[lab=RelTgtStore]
            {
                (\datalangLoc_t + \datalangIdx) \iPointsto_t \datalangVal_t
                \\\\
                (\datalangLoc_t + \datalangIdx) \iPointsto_t \datalangVal_t' \iWand
                \iSimv[\iProt]{\iPred}{\datalangExpr_s}{\datalangUnit}
            }{
                \iSimv[\iProt]{\iPred}{\datalangExpr_s}{\datalangStore{\datalangLoc_t}{\datalangIdx}{\datalangVal_t'}}
            }
        \\
        \inferrule*[lab=RelLoad]
            {
                \datalangLoc_s \iSimilar \datalangLoc_t
            \and
                \forall \datalangVal_s, \datalangVal_t \ldotp
                \datalangVal_s \iSimilar \datalangVal_t
                \iWand
                \iPred (\datalangVal_s, \datalangVal_t)
            }{
                \iSimv[\iProt]{\iPred}{\datalangLoad{\datalangLoc_s}{\datalangIdx}}{\datalangLoad{\datalangLoc_t}{\datalangIdx}}
            }
        \and
        \inferrule*[lab=RelStore]
            {
                \datalangLoc_s \iSimilar \datalangLoc_t
            \and
                \datalangVal_s \iSimilar \datalangVal_t
            \and
                \iPred (\datalangUnit, \datalangUnit)
            }{
                \iSimv[\iProt]{\iPred}{\datalangStore{\datalangLoc_s}{\datalangIdx}{\datalangVal_s}}{\datalangStore{\datalangLoc_t}{\datalangIdx}{\datalangVal_t}}
            }
        \\
        \inferrule*[lab=RelProtocol]
            {
              \iProt (\iPredTwo, \datalangExpr_s, \datalangExpr_t)
            \and
               \forall \datalangExpr_s', \datalangExpr_t' \ldotp
               \iPredTwo (\datalangExpr_s', \datalangExpr_t') \iWand
               \iSimv[\iProt]{\iPred}{\datalangExpr_s'}{\datalangExpr_t'}
            }{
                \iSimv[\iProt]{\iPred}{\datalangExpr_s}{\datalangExpr_t}
            }
    \end{mathparpagebreakable}
    \caption{Relational separation logic (excerpt)}
    \label{fig:program_logic}
\end{figure}

%% file: protocols.tex
\section{Abstract Protocols} \label{sec:protocols} \label{subsec:protocols}

\input{figures/protocol}

In \cref{sec:simulation}, we explain how our relation is defined coinductively and the first step of the proof essentially amounts to coinduction.
To internalize the coinduction hypothesis into the program logic, we introduce an additional parameter $\iProt$, a \emph{protocol}~\citep*{protocols-2021}, which is a general proof-state transformer of type
\begin{mathline}
(\datalangExpr[] \to \datalangExpr[] \to \iProp) \to \datalangExpr[] \to \datalangExpr[] \to \iProp
\end{mathline}

Protocols are used in the logic via the $\RefTirName{RelProtocol}$ rule.
A pair of expressions $\datalangExpr_s$ and $\datalangExpr_t$ is supported by the protocol when it relates them to a postcondition $\iPredTwo$, capturing the possible results of an abstract/axiomatic transition from $\datalangExpr_s$ and $\datalangExpr_t$.
To conclude that $\datalangExpr_s$ and $\datalangExpr_t$ are related, one must prove that any two $\datalangExpr_s'$ and $\datalangExpr_t'$ accepted by this postcondition $\iPredTwo$ remain related.

\subsection{TMC Protocols}

In our correctness proof for the TMC transformation, we use a specific protocol $\iProt_\mathrm{TMC}$ defined in \cref{fig:protocol} by combining two sub-protocols $\iProt_\mathrm{dir}$ and $\iProt_\mathrm{DPS}$ for the direct-style and DPS-style functions. Our coinduction hypothesis assumes toplevel function calls to be compatible with the direct and DPS specifications that we want to prove, and allows to reason about recursive calls to those functions inside the function bodies we are trying to relate.

$\iProt_\mathrm{dir}$ specifies the \emph{direct calling convention} induced by the direct transformation.
It requires $\datalangExpr_s$ and $\datalangExpr_t$ to be function calls to the same function with similar arguments.
To apply it, users can choose any postcondition implied by value similarity.
This rule is equivalent to the \TirName{Sim-Call} rule of \Simuliris.
Most useful protocols are formed by combining $\iProt_\mathrm{dir}$ with other, more specialized protocols.

$\iProt_\mathrm{DPS}$ specifies the \emph{DPS calling convention} induced by the DPS transformation.
It requires $\datalangExpr_s$ to be a function call to a TMC-transformed function $\datalangFn$ and $\datalangExpr_t$ to be a function call to the DPS transform of $\datalangFn$.
As in the DPS specification, ownership of the destination must be passed to the protocol.
To apply it, users can choose any postcondition implied by the postcondition of the DPS specification, including the recovered ownership of the modified destination.

\subsection{Other Examples of Protocols}

Our program logic can be instantiated with other protocols to reason other program transformations.
To demonstrate this generality, we have also verified an inlining and an accumulator-passing-style (APS) transformation --- both included in our mechanization.

\paragraph{Inlining:} Here, the relation $\datalangExpr_s \rightsquigarrow \datalangExpr_t$ allows $\datalangExpr_t$ to recursively inline functions in $\datalangExpr_s$.
As with TMC, it captures all possible inlining strategies.
This relation can be proved correct by using a fairly simple protocol (combined with $\iProt_\mathrm{dir}$) relating a source function and its body:

\begin{tabular}{rcl}
        $\iProt_\mathrm{inline} (\iPredTwo, \datalangExpr_s, \datalangExpr_t)$
        & $\coloneqq$ &
        $\exists \datalangFn, \datalangVar, \datalangExpr_s', \datalangExpr_t', \datalangVal_s, \datalangVal_t \ldotp$
    \\
        &&
        $
        \datalangExpr_s = \datalangCall{\datalangFnptr{\datalangFn}}{\datalangVal_s}
        \ \iSep\ %
        \datalangVal_s \iSimilar \datalangVal_t
        \ \iSep {}$
    \\
        &&
        $\datalangProg_s [\datalangFn] = (\datalangRec{\datalangVar}{\datalangExpr_s'})
        \ \iSep\ %
        \datalangExpr_s' \rightsquigarrow \datalangExpr_t'
        \ \iSep\ %
        \datalangExpr_t = (\datalangLet{\datalangVar}{\datalangVal_t}{\datalangExpr_t'})
        \ \iSep {}$
    \\
        &&
        $\forall \datalangValTwo_s, \datalangValTwo_t \ldotp
        \datalangValTwo_s \iSimilar \datalangValTwo_t \iWand
        \iPredTwo (\datalangValTwo_s, \datalangValTwo_t)$
\end{tabular}
\medskip

\paragraph{Accumulator-passing style}: The APS transformation is a variant of the TMC transformation where the contexts that are made tail-recursive are applications of associative arithmetic operators, typically of the form $(\datalangExpr + \datalangCtxHole)$ or $\datalangExpr_1 + (\datalangExpr_2 \times \datalangCtxHole)$. (See the discussion by \citet*{tmc-koka-2023}.)

We define an APS transformation, after extending \DataLang with integers and arithmetic operations. We verify it with a protocol similar to $\iProt_\mathrm{DPS}$ that allows calling the APS transform of a source function with an integer accumulator: if $\datalangCall{\datalangFn}{\datalangVal_s}$ returns $n$, then $\datalangCall{\datalangFn_\mathit{aps}}{\datalangPair{\datalangVal_{\mathit{acc}}}{\datalangVal_t}}$ returns $\datalangVal_\mathit{acc} + n$.

\medskip
\begin{tabular}{rcl}
        $\iProt_\mathrm{APS} (\iPredTwo, \datalangExpr_s, \datalangExpr_t)$
        & $\coloneqq$ &
        $\exists \datalangFn, \datalangFn_\mathit{aps}, \datalangVal_s, \datalangVal_\mathit{acc}, \datalangVal_t \ldotp$
    \\
        &&
        $\datalangFn \in \dom{\datalangProg_s}
        \ \iSep\ %
        \datalangRenaming [\datalangFn] = \datalangFn_\mathit{aps}
        \iSep {}$
    \\
        &&
        $\datalangVal_s \iSimilar \datalangVal_t
        \ \iSep\ %
        \datalangExpr_s = \datalangCall{\datalangFnptr{\datalangFn}}{\datalangVal_s}
        \ \iSep\ %
        \datalangExpr_t = \datalangCall{\datalangFnptr{\datalangFn_\mathit{aps}}}{\datalangPair{\datalangVal_\mathit{acc}}{\datalangVal_t}}
        \iSep {}$
    \\
        &&
        $\forall \datalangVal_s', \datalangExpr_t' \ldotp$
    \\
        &&
        $\bm{\mathrm{match}}\ \datalangVal_s'\ \bm{\mathrm{with}}\ \datalangNat \Rightarrow \datalangExpr_t' = \datalangVal_\mathit{acc} + \datalangNat \mid \_ \Rightarrow \mathrm{strongly \mathhyphen stuck}_{\datalangProg_t} (\datalangExpr_t')\ \bm{\mathrm{end}} \iWand$
    \\
        &&
        $\iPredTwo (\datalangVal_s', \datalangExpr_t')$
\end{tabular}
\medskip
\Xfrancois{Explain why this guarantee (``$f_\mathit{aps}$ will get stuck'') is needed. Do we want to prove that if the source crashes then the target crashes?}

One subtlety is that our \DataLang language is untyped, so arithmetic operations (here addition) may get stuck on non-integer values. If the function call $\datalangCall {\datalangFnptr \datalangFn} {\datalangVal_s}$ in the source program returns a non-integer value, then the outer context $\datalangVal_\mathit{acc} + \datalangCtxHole$ gets stuck. But\Xfrancois{not easy to follow your reasoning here} in the transformed program, this failure happens inside the body of the APS-transformed function $\datalangFnptr {\datalangFn_\mathit{aps}}$. To represent this failure case in our protocol, the postcondition $\iPredTwo$ relates a non-integer source return value $\datalangVal_s'$ with any strongly stuck expression $\datalangExpr_t'$ in the target. This relies on the generality of our protocols being predicate transformers on expressions, not just values.


%% file: figures/protocol.tex
\begin{figure}[tp]
    \begin{tabular}{rcl}
            $\iProt_\mathrm{dir} (\iPredTwo, \datalangExpr_s, \datalangExpr_t)$
            & $\coloneqq$ &
            $\exists \datalangFn, \datalangVal_s, \datalangVal_t \ldotp$
        \\
            &&
            $\datalangFn \in \dom{\datalangProg_s} \iSep
            \datalangExpr_s = \datalangCall{\datalangFnptr{\datalangFn}}{\datalangVal_s} \iSep
            \datalangExpr_t = \datalangCall{\datalangFnptr{\datalangFn}}{\datalangVal_t} \iSep {}$
        \\
            &&
            $\datalangVal_s \iSimilar \datalangVal_t \iSep {}$
        \\
            &&
            $\forall \datalangValTwo_s, \datalangValTwo_t \ldotp
            \datalangValTwo_s \iSimilar \datalangValTwo_t \iWand
            \iPredTwo (\datalangValTwo_s, \datalangValTwo_t)$
        \\
            $\iProt_\mathrm{DPS} (\iPredTwo, \datalangExpr_s, \datalangExpr_t)$
            & $\coloneqq$ &
            $\exists \datalangFn, \datalangFn_\mathit{dps}, \datalangVal_s, \datalangLoc_1, \datalangLoc_2, \datalangLoc, \datalangIdx, \datalangVal_t \ldotp$
        \\
            &&
            $\datalangFn \in \dom{\datalangProg_s} \iSep
            \datalangRenaming [\datalangFn] = \datalangFn_\mathit{dps} \iSep
            \datalangExpr_s = \datalangCall{\datalangFnptr{\datalangFn}}{\datalangVal_s} \iSep
            \datalangExpr_t = \datalangCall{\datalangFnptr{\datalangFn_\mathit{dps}}}{\datalangLoc_1} \iSep {}$
        \\
            &&
            $(\datalangLoc_1 + 1) \iPointsto_t (\datalangLoc_2, \datalangVal_t) \iSep
            (\datalangLoc_2 + 1) \iPointsto_t (\datalangLoc, \datalangIdx) \iSep
            (\datalangLoc + \datalangIdx) \iPointsto \datalangHole \iSep
            \datalangVal_s \iSimilar \datalangVal_t$
        \\
            &&
            $\forall \datalangValTwo_s, \datalangValTwo_t \ldotp
            (\datalangLoc + \datalangIdx) \iPointsto \datalangValTwo_t \iSep
            \datalangValTwo_s \iSimilar \datalangValTwo_t \iWand
            \iPredTwo (\datalangValTwo_s, \datalangUnit)$
        \\
            $\iProt_\mathrm{TMC}$
            & $\coloneqq$ &
            $\iProt_\mathrm{dir} \sqcup \iProt_\mathrm{DPS}
            =
            \lambdaAbs (\iPredTwo, \datalangExpr_s, \datalangExpr_t) \ldotp \iProt_\mathrm{dir} (\iPredTwo, \datalangExpr_s, \datalangExpr_t) \iOr \iProt_\mathrm{DPS} (\iPredTwo, \datalangExpr_s, \datalangExpr_t)$
    \end{tabular}
    \caption{TMC protocol ($\iProt_\mathrm{TMC}$)}
    \label{fig:protocol}
\end{figure}

%% file: proof.tex
\section{Proof of the Specification}
\label{sec:proof}

In this section, we prove the specifications of \cref{sec:specification}.

As mentioned in \cref{sec:protocols}, we instantiate our program logic with a specific protocol $\iProt_\mathrm{TMC}$ defined in \cref{fig:protocol}. We define a shorthand notation for this instantiation:
\begin{mathline}
    \iSimv{
        \iPred
    }{
        \datalangExpr_s
    }{
        \datalangExpr_t
    }
    \coloneqq
    \iSimv[
        \iProt_\mathrm{TMC}
    ]{
        \iPred
    }{
        \datalangExpr_s
    }{
        \datalangExpr_t
    }
\end{mathline}

So far, we worked with \emph{closed} expressions, that have no free variables.\Xfrancois{This was never said.}
We need to generalize the specifications to \emph{open} expressions that may have free variables, as is standard.
To do so, we introduce a \emph{runtime relation} $\iRsimv{\iPred}{\datalangExpr_s}{\datalangExpr_t}$ in \cref{fig:rsim}.
It requires $\datalangBisubst_s (\datalangExpr_s)$ and $\datalangBisubst_t (\datalangExpr_t)$ to be related for any \emph{well-formed closing bisubstitution} $\datalangBisubst \in \datalangVar[] \rightarrow \datalangVal[] \times \datalangVal[]$.
In practice, $\datalangBisubst$ contains $\datalangLetKeyword$-bound variables that have been $\beta$-reduced, and their substitute source and target values.

In addition, we will only consider \emph{valid source expressions} --- denoted by $\wf{\datalangExpr_s}$ ---, \ie those that do not involve any location, deterministic block expressions, or undefined source function.

\begin{lemma}[Specification of direct transformation] \label{thm:dir}
    \[
        \iRsimvHoare{
            \wf{\datalangExpr_s} \iSep
            \datalangExpr_s \tmcDir{\datalangRenaming} \datalangExpr_t
        }{
            \datalangVal_s, \datalangVal_t \ldotp
            \datalangVal_s \iSimilar \datalangVal_t
        }{
            \datalangExpr_s
        }{
            \datalangExpr_t
        }
    \]
\end{lemma}
\Xfrancois{Here and below the pure assumptions should be put outside, with an implication, they are ``static'' (compile-time) assumptions.}

\begin{lemma}[Specification of DPS transformation] \label{thm:dps}
    \[
        \iRsimvHoare{
            \wf{\datalangExpr_s} \iSep
            (\datalangLoc, \datalangIdx, \datalangExpr_s) \tmcDps{\datalangRenaming} \datalangExpr_t \iSep
            (\datalangLoc + \datalangIdx) \iPointsto_t \datalangHole
        }{
            \datalangVal_s, \datalangUnit \ldotp
            \exists \datalangVal_t \ldotp
            (\datalangLoc + \datalangIdx) \iPointsto_t \datalangVal_t \iSep
            \datalangVal_s \iSimilar \datalangVal_t
        }{
            \datalangExpr_s
        }{
            \datalangExpr_t
        }
    \]
\end{lemma}
\Xfrancois{TODO point back to the beginning of the paper for an informal explanation of these statements.}

Both proofs proceed by induction over $\datalangExpr_s$ and mutual induction over $\datalangExpr_s \tmcDir{\datalangRenaming} \datalangExpr_t$ and $(\datalangLoc, \datalangIdx, \datalangExpr_s) \tmcDps{\datalangRenaming} \datalangExpr_t$. In each case we then apply the relevant rules of the program logic. 

\input{figures/rsim}


%% file: figures/rsim.tex
\begin{figure}[tp]
    \begin{align*}
        	\wf{\datalangBisubst}
    		& \coloneqq
    		\forall \datalangVar \ldotp
    		\exists \datalangVal_s, \datalangVal_t \ldotp
    		\datalangBisubst (\datalangVar) = (\datalangVal_s, \datalangVal_t) \iSep
    		\datalangVal_s \iSimilar \datalangVal_t
        \\
    		\iRsimv{\iPred}{\datalangExpr_s}{\datalangExpr_t}
    		& \coloneqq
    		\forall \datalangBisubst \ldotp
    		\wf{\datalangBisubst} \iWand
    		\iSimv{\iPred}{{\datalangBisubst (\datalangExpr_s)}_1}{{\datalangBisubst (\datalangExpr_t)}_2}
    	\\
    	   \iRsimvHoare{P}{\iPred}{\datalangExpr_s}{\datalangExpr_t}
    	   & \coloneqq
    	   \iPersistent \left( P \iWand \iRsimv{\iPred}{\datalangExpr_s}{\datalangExpr_t} \right)
    \end{align*}
    \caption{Runtime relation}
    \label{fig:rsim}
\end{figure}

%% file: simulation.tex
\section{Simulation}
\label{sec:simulation}

So far, we assumed a program logic satisfying a set of reasoning rules.
In this section, we prove that our rules are sound: they imply a \emph{simulation} à la \Simuliris~\citep*{simuliris-2022}.
This simulation comes with an \emph{adequacy theorem} that allows to extract a \emph{behavioral refinement} in the meta-logic (Coq, without \Iris), our final soundness theorem.

\subsection{Definition}

\input{figures/sim}
\input{figures/refinement}

Our simulation in defined in \cref{fig:sim}.
It is largely inspired by the \Simuliris simulation --- simplified due to the absence of concurrency.
The main difference lies in the protocol clause \circled{5}, which is a generalization of the function call clause of \Simuliris.

\newcommand{\iSimGfp}{\textcolor{\iSimGfpColor}{sim}\xspace}
\newcommand{\iSimLfp}{\textcolor{\iSimLfpColor}{sim-inner}\xspace}

The definition consists of two nested fixpoints \iSimGfp and \iSimLfp.
\iSimGfp is a greatest fixpoint (coinduction) allowing expressions to diverge (in a controlled way, see clause \circled{4}\circled{B}).
\iSimLfp is a least fixpoint (induction) allowing source and target stuttering (see clauses \circled{3} and \circled{4}\circled{A}).
The \emph{state interpretation} $\mathrm{I} (\datalangState_s, \datalangState_t)$ intuitively materializes the invariant of the simulation, including the heap bijection (see \cref{sec:specification}); it is systematically maintained.
We now review the six clauses.

\paragraph{\circled{1} Postcondition.}
The simulation can stop when the postcondition is satisfied.

\paragraph{\circled{2} Stuck expressions.}
The simulation can also stop on simultaneously stuck expressions.

\paragraph{\circled{3} Target stuttering.}
The source expression can angelically take some steps.
\Xfrancois{Could this be just one step?}
\Xgabriel{I think the present form allows to temporarily break the state invariant.}
This can only happen finitely many times, as we continue with \iSimLfp.
If we used \iSimGfp instead, a silent loop in the source could be simulated by anything, breaking preservation of divergence.

\paragraph{\circled{4}\circled{A} Source stuttering.}
The target expression can demonically take one step.
This can also only happen finitely many times, as we continue with \iSimLfp.
If we used \iSimGfp instead, a silent loop in the target could simulate any source expression, breaking preservation of termination.

\paragraph{\circled{4}\circled{B} Synchronization.}
Alternatively, both expressions can simultaneously take one step.
This can happen infinitely many times, as we continue with \iSimGfp.
If we used \iSimLfp instead, we would be unable to relate divergent programs.

\paragraph{\circled{5} Protocol application.}
Finally, we can apply the protocol under evaluation contexts.
We can choose any postcondition $\iPredTwo$ accepted by the protocol and assume it to prove the continuation.
(We justify separately that a protocol is admissible, see \cref{subsec:closure}.)

\subsection{Simulation Closure}\label{subsec:closure}

If a protocol $\iProt$ respects a certain admissibility condition, then program relations established using this protocol are also in the \emph{closed} simulation, using the empty protocol $\bot$.

\begin{definition}[Admissibility]
  \label{def:protocol-admissibility}
  A protocol $\iProt$ is admissible, written $\Admissible{\iProt}$, when we have:
  \begin{mathline}
    \iPersistent \left(
      \forall \iPredTwo, \datalangExpr_s, \datalangExpr_t \ldotp
      \iProt (\iPredTwo, \datalangExpr_s, \datalangExpr_t) \iWand
      \mathrm{sim \mathhyphen inner}_\bot (\lambdaAbs (\_, \datalangExpr_s', \datalangExpr_t') \ldotp \iSim[\iProt]{\iPredTwo}{\datalangExpr_s'}{\datalangExpr_t'}) (\bot, \datalangExpr_s, \datalangExpr_t)
    \right)
  \end{mathline}
\end{definition}

In simple terms, the admissibility condition $\Admissible{\iProt}$ states that every triple $(\iPredTwo, \datalangExpr_s, \datalangExpr_t)$ is justified, that is, that $\datalangExpr_s$ and $\datalangExpr_t$ are related.
But the protocol $\iProt$ cannot be used right away to establish this relation (this would allow cyclic, vacuous proofs). Our use of $\mathrm{sim \mathhyphen inner}_\bot$ forces a proof of admissibility to perform some ``productive'' simulation steps with an empty protocol: in this instantiation of the simulation, \iSimLfp uses the empty protocol and \iSimGfp uses $\iProt$, so we have to perform at least one reduction in the source before we can use the protocol again.

\begin{theorem}[Simulation closure]
\label{thm:closure}
  For any protocol $\iProt$, we have:
    \begin{mathline}
            \Admissible{\iProt} \iWand
            \iSim[\iProt]{\iPred}{\datalangExpr_s}{\datalangExpr_t} \iWand
            \iSim[\bot]{\iPred}{\datalangExpr_s}{\datalangExpr_t}
    \end{mathline}
\end{theorem}

Actually, for the TMC protocol, we prove a simpler condition that implies admissibility:
\begin{mathline}
            \iPersistent \left(
                \forall \iPredTwo, \datalangExpr_s, \datalangExpr_t \ldotp
                \iProt (\iPredTwo, \datalangExpr_s, \datalangExpr_t) \iWand
                \exists \datalangExpr_s', \datalangExpr_t' \ldotp
                \datalangExpr_s \pureStep{\datalangProg_s} \datalangExpr_s' \iSep
                \datalangExpr_t \pureStep{\datalangProg_t} \datalangExpr_t' \iSep
                \iSim[\iProt]{\iPredTwo}{\datalangExpr_s'}{\datalangExpr_t'}
            \right)
\end{mathline}
With this weaker version, an admissibility proof must perform exactly one pure step on both sides before the protocol $\iProt$ becomes available again. Other users of our program logic may want to reuse this simpler definition, unless they need the full generality of the $\Admissible{\iProt}$ definition.

\subsection{Adequacy}

Informally, our closed simulation is \emph{adequate} in the sense that if $\datalangExpr_s$ simulates $\datalangExpr_t$, then $\datalangExpr_t$ refines $\datalangExpr_s$, \ie the behaviors of $\datalangExpr_t$ are included in the behaviors of $\datalangExpr_s$:

\begin{theorem}[Simulation adequacy] \label{thm:adequacy}
    $
        \left( \vdash \iSimv[\bot]{\iSimilar}{\datalangExpr_s}{\datalangExpr_t} \right) \implies
        \datalangExpr_s \refined \datalangExpr_t
    $
\end{theorem}

The notions of \emph{behaviors} and \emph{refinement} are defined in \cref{fig:refinement}.
We consider not only converging behaviors (resulting in values or stuck expressions) but also diverging behaviors.
Expression refinement $\datalangExpr_s \refined \datalangExpr_t$ is \emph{termination-preserving}: if $\datalangExpr_s$ always terminates, so does $\datalangExpr_t$.
Program refinement $\datalangProg_s \refined \datalangProg_t$, also defined in \cref{fig:refinement}, requires any source function call in $\datalangProg_t$ to behave as in $\datalangProg_s$.\Xfrancois{ambiguous}

\subsection{Transformation Soundness}

We can finally express the soundness of the TMC transformation: if program $\datalangProg_s$ is well-formed and transforms into program $\datalangProg_t$, then $\datalangProg_t$ refines $\datalangProg_s$.
A program is well-formed when its function definitions are well-formed and well-scoped.
Note that this statement does not use separation logic. (In our mechanization, it is a pure Coq statement without Iris propositions.)

\begin{theorem}[Transformation soundness] \label{thm:soundness}
    $
        \wf{\datalangProg_s} \wedge \datalangProg_s \tmc \datalangProg_t \implies
        \datalangProg_s \refined \datalangProg_t
    $
\end{theorem}



%% file: figures/sim.tex
\begin{figure}[tp]
    \begin{align*}
    		\mathrm{sim \mathhyphen body}_\iProt
    		&\coloneqq
    		\begin{array}{l}
    				\lambdaAbs \textcolor{\iSimGfpColor}{sim} \ldotp
    				\lambdaAbs \textcolor{\iSimLfpColor}{sim \mathhyphen inner} \ldotp
    				\lambdaAbs {(\iPred, \datalangExpr_s, \datalangExpr_t)} \ldotp
    				\forall \datalangState_s, \datalangState_s \ldotp
    				\mathrm{I} (\datalangState_s, \datalangState_t)
    				\iWand \iBupd
    			\\
    				\bigvee \left[ \begin{array}{ll}
    							\circled{1}
    						&
    							\mathrm{I} (\datalangState_s, \datalangState_t) \iSep
    							\iPred (\datalangExpr_s, \datalangExpr_t)
    					\\
    					        \circled{2}
                            &
                                \mathrm{I} (\datalangState_s, \datalangState_t) \iSep
    							\mathrm{strongly \mathhyphen stuck}_{\datalangProg_s} (\datalangExpr_s) \iSep
    							\mathrm{strongly \mathhyphen stuck}_{\datalangProg_t} (\datalangExpr_s)
    					\\
    							\circled{3}
    						&
    							\exists \datalangExpr_s', \datalangState_s' \ldotp
    							(\datalangExpr_s, \datalangState_s) \step{\datalangProg_s}^+ (\datalangExpr_s', \datalangState_s') \iSep
    							\mathrm{I} (\datalangState_s', \datalangState_t) \iSep
    							\textcolor{\iSimLfpColor}{sim \mathhyphen inner} (\iPred, \datalangExpr_s', \datalangExpr_t)
    					\\
    							\circled{4}
    						&
								\mathrm{reducible}_{\datalangProg_t} (\datalangExpr_t, \datalangState_t) \iSep
								\forall \datalangExpr_t', \datalangState_t' \ldotp
								(\datalangExpr_t, \datalangState_t) \step{\datalangProg_t} (\datalangExpr_t', \datalangState_t')
								\iWand \iBupd
						\\
                            &
								\bigvee \left[ \begin{array}{ll}
											\circled{A}
										&
											\mathrm{I} (\datalangState_s, \datalangState_t') \iSep
											\textcolor{\iSimLfpColor}{sim \mathhyphen inner} (\iPred, \datalangExpr_s, \datalangExpr_t')
									\\
											\circled{B}
										&
											\exists \datalangExpr_s', \datalangState_s' \ldotp
											(\datalangExpr_s, \datalangState_s) \step{\datalangProg_s}^+ (\datalangExpr_s', \datalangState_s') \iSep
											\mathrm{I} (\datalangState_s', \datalangState_t') \iSep
											\textcolor{\iSimGfpColor}{sim} (\iPred, \datalangExpr_s', \datalangExpr_t')
								\end{array} \right.
    					\\
    							\circled{5}
    						&
    							\exists \datalangEctx_s, \datalangExpr_s', \datalangEctx_t, \datalangExpr_t', \iPredTwo \ldotp
    					\\
    					    &
    					        \datalangExpr_s = \datalangEctx_s [\datalangExpr_s'] \iSep
    							\datalangExpr_t = \datalangEctx_t [\datalangExpr_t'] \iSep
    						    \iProt (\iPredTwo, \datalangExpr_s', \datalangExpr_t') \iSep
    						    \mathrm{I} (\datalangState_s, \datalangState_t) \iSep {}
    					\\
                            &
								\forall \datalangExpr_s'', \datalangExpr_t'' \ldotp
								\iPredTwo (\datalangExpr_s'', \datalangExpr_t'') \iWand
								\textcolor{\iSimLfpColor}{sim \mathhyphen inner} (\iPred, \datalangEctx_s [\datalangExpr_s''], \datalangEctx_t [\datalangExpr_t''])
    				\end{array} \right.
    		\end{array}
    	\\
    	    \mathrm{\color{\iSimLfpColor}{sim \mathhyphen inner}}_\iProt
    	    &\coloneqq
    	    \lambdaAbs \textcolor{\iSimGfpColor}{sim} \ldotp
    	    \muAbs \textcolor{\iSimLfpColor}{sim \mathhyphen inner} \ldotp
    	    \mathrm{sim \mathhyphen body}_\iProt (sim, \textcolor{\iSimLfpColor}{sim \mathhyphen inner})
    	\\
    		\mathrm{\color{\iSimGfpColor}{sim}}_\iProt
    		&\coloneqq
    		\nuAbs \textcolor{\iSimGfpColor}{sim} \ldotp
    		\mathrm{\color{\iSimLfpColor}{sim \mathhyphen inner}}_\iProt (\textcolor{\iSimGfpColor}{sim})
    	\\
    		\iSim[\iProt]{\iPred}{\datalangExpr_s}{\datalangExpr_t}
    		&\coloneqq
    		\mathrm{\color{\iSimGfpColor}{sim}}_\iProt (\iPred, \datalangExpr_s, \datalangExpr_t)
    	\\
    	   \iSimv[\iProt]{\iPred}{\datalangExpr_s}{\datalangExpr_t}
    	   &\coloneqq
    	   \iSim[\iProt]{\lambdaAbs (\datalangExpr_s', \datalangExpr_t') \ldotp \exists \datalangVal_s, \datalangVal_t \ldotp \datalangExpr_s' = \datalangVal_s \iSep \datalangExpr_t' = \datalangVal_t \iSep \iPred (\datalangVal_s, \datalangVal_t)}{\datalangExpr_s}{\datalangExpr_t}
    \end{align*}
    \caption{Simulation with protocol}
    \label{fig:sim}
\end{figure}

%% file: figures/refinement.tex
\begin{figure}[tp]
    \centering
    \begin{mathparpagebreakable}
        \inferrule*
            {}{
                \datalangUnit \similar \datalangUnit
            }
        \and
        \inferrule*
            {}{
                \datalangIdx \similar \datalangIdx
            }
        \and
        \inferrule*
            {}{
                \datalangTag \similar \datalangTag
            }
        \and
        \inferrule*
            {}{
                \datalangBool \similar \datalangBool
            }
        \and
        \inferrule*
            {}{
                \datalangLoc_s \similar \datalangLoc_t
            }
        \and
        \inferrule*
            {}{
                \datalangFnptr{\datalangFn} \similar \datalangFnptr{\datalangFn}
            }
        \\
        \inferrule*
            {
                \datalangVal_s \similar \datalangVal_t
            }{
                \constr[\datalangVal_s]{Conv} \refined \constr[\datalangVal_t]{Conv}
            }
        \and
        \inferrule*
            {
                \datalangExpr_s \notin \datalangVal[]
            \and
                \datalangExpr_t \notin \datalangVal[]
            }{
                \constr[\datalangExpr_s]{Conv} \refined \constr[\datalangExpr_t]{Conv}
            }
        \and
        \inferrule*
            {}{
                \constr{Div} \refined \constr{Div}
            }
    \end{mathparpagebreakable}
    \\
    \begin{tabular}{rcl}
        	$\mathrm{behaviours}_{\datalangProg} (\datalangExpr)$
        	& $\coloneqq$ &
			$\{ \constr[\datalangExpr']{Conv} \mid \exists \datalangState \ldotp (\datalangExpr, \emptyset) \step{\datalangProg}^* (\datalangExpr', \datalangState) \wedge \mathrm{irreducible}_{\datalangProg} (\datalangExpr', \datalangState) \}\ \uplus$
    	\\
            &&
    		$\{ \constr{Div} \mid\ \diverges{\datalangProg}{(\datalangExpr, \emptyset)} \}$
        \\
            $\datalangExpr_s \refined \datalangExpr_t$
            & $\coloneqq$ &
            $\forall b_t \in \mathrm{behaviours}_{\datalangProg_t} (\datalangExpr_t) \ldotp
            \exists b_s \in \mathrm{behaviours}_{\datalangProg_s} (\datalangExpr_s) \ldotp
            b_s \refined b_t$
        \\
            $\datalangProg_s \refined \datalangProg_t$
            & $\coloneqq$ &
            $\forall \datalangFn \in \dom{\datalangProg_s}, \datalangVal \ldotp
            \wf{\datalangVal} \implies
            \datalangCall{\datalangFnptr{\datalangFn}}{\datalangVal} \refined \datalangCall{\datalangFnptr{\datalangFn}}{\datalangVal}$
    \end{tabular}
    \caption{Program refinement}
    \label{fig:refinement}
\end{figure}

%% file: related.tex
\section{Related Work}

\subsection{TMC Support in Compilers}

Tail-recursion modulo cons was well-known in the Lisp community as
early as the 1970s. For example the REMREC system~\citep*{risch-73}
automatically transforms recursive functions into loops, and
supports modulo-cons tail recursion. It also supports tail-recursion
modulo associative arithmetic operators, which is outside the scope
of our work, but supported by the GCC compiler for example. The TMC
fragment is precisely described (in prose) by \citet*{friedman-wise-75}. Other implementations include OPAL~\citep*{opal-1994}.

In the Prolog community it is a common pattern to implement destination-passing style through unification variables; in particular ``difference lists'' are a common representation of lists with a final hole.
Unification variables are first-class values: in particular they can be passed as function arguments, providing expressive, first-class support for destination-passing style in the source language.
For example, we do not support optimizing tail contexts of the form \ocaml{List.append li _?}, only direct constructor applications; this can be expressed in Prolog as just the difference list \ocaml{(List.append li X, X)} for a fresh destination variable \ocaml{X}.
But this expressiveness comes at a performance cost, and there is no static checking that the data is fully initialized at the end of computation.

\citet*{tmc-scala-2013} implement Ozma, an experimental back-end for
the Scala programming language that targets the Oz virtual machine.
Oz~\citep*{oz-1994,oz-1995} is a language that integrates features
from logic- and functional-programming style, and in particular offers
pervasive ``dataflow values'', a generalization of Prolog unification
variables. Ozma brings to Scala an idiom of Oz, where Prolog-style
difference lists are used to represent potentially-infinite streams
that model synchronous concurrent agents. These lists must be
processed in constant stack space, so Ozma introduces
a tail-modulo-cons transformation in Prolog fashion: all data
constructor arguments (and some explicitly-annotated
function arguments) are transformed into dataflow values. The
motivation is expressiveness, not performance, and the Ozma back-end is
not competitive with the Scala JVM back-end. Besides the obvious
engineering-effort differences, the pervasive use of dataflow values
may incur high constant overhead, making this approach unsuitable to
bring TMC to performance-conscious Scala users.

Independently of our work, Koka has implemented TMC starting in August
2020\footnote{\url{https://github.com/koka-lang/koka/commit/f6a343d31f486ea5edd44798dca7bca52d7b450c}}~\citep*{tmc-koka-2023}.
An interesting problem they had to solve, which does not occur in \OCaml,
is how to support TMC in presence of non-linear continuations. Our
correctness argument for TMC relies on the fact that the destination
is uniquely owned, and written exactly once; this property may not
hold in programs that use multishot continuations (\ocaml|call/cc|,
\ocaml|let/cc|, \ocaml|delim/cc|) or multishot effect handlers. The standard Koka runtime uses its
reference-counting machinery to determine that a destination is not
uniquely-owned anymore, and stores extra metadata in
partially-initialized blocks to be able to copy them on-demand in this
case. Its JavaScript back-end instead reverts to a CPS transformation
when non-linear control flow is detected.

\subsection{Reasoning About Destination-Passing-Style}

In general, if we think of non-tail recursive functions as having an ``evaluation context'' representing the continuation of the recursive call, then the techniques to turn classes of calls into tail-calls correspond to different reified representations of non-tail contexts, equipped with specific (efficient) implementations of context composition and hole-plugging.
TMC comes from representing data-construction contexts as the partial data itself, with hole-plugging by mutation.
Associative-operator transformations represent the context \ocaml|1 + (4 + _?)| as the number \ocaml|5| directly.
(Sometimes it suffices to keep around an abstraction of the context; see John Clements' work on stack-based security~\citep*{clements-2004}.)

\citet*{minamide-98} gives a ``functional'' interface to destination-passing-style programs, by presenting a partial data-constructor composition \ocaml{Foo(x,Bar(_?))} as a use-once, linear-typed function \ocaml{linfun h -> Foo(x,Bar(h))}.
Those special linear functions remain implemented as partial data, but they expose a referentially-transparent interface to the programmer, restricted by a linear type discipline.
This is a beautiful way to represent destination-passing style, orthogonal to our work: users of Minamide's system would still have to write the transformed version by hand, and we could implement a transformation into destination-passing style expressed in his system.
\citet*{sobel-friedman-98}, inspired by Minamide's work, optimize continuation-passing-style versions of tree-traversal functions: they defunctionalize the continuations and systematically derive a pointer-inversion implementation that remains tail-recursive. \citet*{bagrel-2023} expresses destination-passing style programming in Linear Haskell. Finally, \citet*{constructor-contexts-2024} propose \emph{constructor contexts} as a first-class data structure corresponding to Minamide's constructor continuations, which results in a more declarative style than traditional DPS programs, yet more explicit and more expressive than just the tail-modulo-cons fragment. They show that traditional imperative tree-traversal programs can be systematically reconstructed from functional implementations via constructor contexts, furthering the relations suggested by \citet*{sobel-friedman-98}.

Separation logic can also model partial structures to be filled later through the \emph{magic wand}, notably used to reason about list segments in imperative list-traversal functions~\citep*{}. Mezzo~\citep*{mezzo-2016} provides a general-purpose type system based on separation logic, which can directly express uniquely-owned partially-initialized data, and its transformation into immutable, duplicable results. (See the \href{https://protz.github.io/mezzo/code_samples/list.mz.html}{List} module of the Mezzo standard library, and in particular \ocaml{cell}, \ocaml{freeze} and \ocaml{append} in destination-passing-style).

\subsection{Correctness Proof for TMC}

\citet*{tmc-koka-2023} provide a pen-and-paper correctness argument for TMC, or in fact a family of approaches based on optimized representations of classes of non-tail contexts, in the style of program calculation.
The clarity of their exposition is remarkable.

We were inspired by the generality of their presentation and verified that our proof technique can also be applied to some other TMC variants, by extending our mechanized development with a correctness proof for an
accumulator-passing-style transformation.

Finally, our correctness results are more precise and slightly stronger: they work in a well-typed setting where programs do not fail, whereas we use untyped terms and show preservation of failure; their proofs assume a deterministic, non-effectful language, whereas we use a more general non-deterministic, effectful language; finally, they have a very simple definition of program equivalence (reducing to the exact same value) that works well for semi-formal reasoning, but is unsuitable to scale the argument to other programming languages, whereas we use a standard notion of behavioral refinement that can scale to less idealized settings.
We get this extra generality mostly as a direct result of our methodology (relational program logic backed by a simulation); but showing preservation of failure requires some care when handling arithmetic operators in the accumulator-passing-style variant.

Remark: the implementation they prove correct, which corresponds to the approach described in \citet*{minamide-98}, results in a slightly less efficient generation where recursive calls to \ocaml{map_dps} are passed \emph{both} the start of the list and the destination to be written at its end. In our case the start of the list remains constant over all recursive calls, so our DPS version does not propagate it. They cannot perform this simplification due to Koka's support for multishot continuations, which sometimes require copying the partial list within the recursion -- the start of the list is necessary at in this case.

\subsection{Relational Reasoning in Separation Logic}

Defining a program logic to capture unary program properties is a typical usage of \Iris; relational properties are rarer. \citet*{tassarotti-2017} use (a linear variant of) \Iris to prove the correctness of a program transformation that implements communication channels using shared references. See the related work of ReLoC Reloaded~\citep*{reloc-2021}.
\Xfrancois{+ thèse de Friis Vindum.}
\Xfrancois{ReLoC a été utilisé surtout pour vérifier des algos concurrents.}

A relational program logic can be justified in \Iris by interpreting it as a unary relation on the target program, typically involving the \texttt{wp} predicate of the base language. This approach is inspired by CaReSL~\citep*{caresl-2013}. We follow a more direct, traditional approach of interpreting the program logic as a (binary) simulation relation (defined in the \Iris meta-logic) which is shown (adequacy) to imply a refinement between the program behaviors (denotations).

It is in fact surprisingly difficult to define simulations in \Iris, if we expect them to be adequate (to correspond to the usual notion of simulation outside the \Iris world). This is due to meta-theoretical difficulties around the ``later'' modality which led to the Transfinite \Iris variant~\citep*{transfinite-iris-2021}. We started by defining simulations in Transfinite \Iris, but later moved to the \Simuliris approach~\citep*{simuliris-2022}, where simulations are defined in standard \Iris without using the ``later'' modality, using coinduction instead (via its impredicative encoding).

As a minor technical point of comparison to \Simuliris, our definition of behaviors (denotations) includes non-termination, successfully evaluating to a value, but also failing with an error, and refinement preserves all three kind of behaviors. We do not model undefined behaviors.

We believe that our approach (relational program logics justified by a simulation) is showing promises for compiler verification. Verification of CompCert\Xgabriel{TODO citation} passes typically prove a forward simulation result, which is strengthened into a backward simulation thanks to a determinism assumption. We get the desired backward simulation directly, with compositional proofs.

\subsection{Protocols}

The function-call rule of \Simuliris only relates calls to the same function, so it is unsuitable for program transformations that also transform function definitions. We parameterize our program logic and notion of simulation on a \emph{protocol} $\iProt$, an arbitrary predicate transformer injected into the relation. This approach is reminiscent of the \emph{axiomatic semantics}~\citep*{axiomatic-semantics-2014} proposed to reason about foreign function calls. In the \Iris community, we were directly inspired by \emph{protocols} \citet*{protocols-2021}, and this approach was also reused recently, in a unary setting, by \citet*{melocoton-2023}. Our notion of protocol is slightly more general than in those two works, as it can relate arbitrary expressions in evaluation position (not just function calls), and ``return'' after an axiomatic transition with arbitrary expressions (not just values). We use this extra generality to reason about accumulator-passing-style transformation in presence of ill-typed programs -- see \cref{subsec:protocols}.


%% file: acknowledgments.tex
\begin{acks}
  This work was initiated by Frédéric Bour who implemented TMC in an
  experimental variant of the OCaml compiler and, in 2015, submitted his work
  for inclusion. At the time, the proposal remained stalled due to lack
  of review and integration effort among maintainers. The broad lines
  of the final implementation, in term of generated code, were already
  in Frédéric Bour's initial version, as well as the idea to have an
  opt-in transformation controlled by an attribute
  (\Appendices{\cref{subsec:optin}}{See Appendix~A.2.1}). This experiment also generated
  performance data that motivated us to push further. (One notable
  technical difference is that instead of letting a destination-passing-style
  function take two parameters \ocaml|dst| and \ocaml|ofs|, Bour
  would generate several versions of the function,
  where the parameter \ocaml|ofs| was specialized to
  a constant. He found that this did not noticeably improve
  performance, and changed back to a parameter to simplify the
  implementation and reduce code size.)

  In 2020, Gabriel Scherer restarted Frédéric Bour's effort with
  a review followed by a partial re-implementation of the
  transformation that introduced the applicative style discussed in
  \Appendices{\cref{subsec:applicative-impl}}{Appendix~A.4} as well as
  much of the current attribute-based user interface to control the
  transformation
  (\Appendices{\cref{subsec:user-control}}{Appendix~A.2.2}). Basile
  Clément in turn reviewed Scherer's version, and introduced
  constructor compression
  (\cref{subsec:constructor-compression}). Xavier Leroy implemented
  a change to the \OCaml calling convention to remove parameter-number
  restrictions on tail calls on some architectures
  (\Appendices{\cref{subsec:PR-history}}{Appendix~A.3}). The work was
  finally reviewed by Pierre Chambart and merged in the upstream
  \OCaml compiler in November 2021.

  Clément Allain started working on a mechanized soundness proof for
  the TMC transformation in Iris in Summer 2022, as a master's
  internship supervised by François Pottier. They discovered that the
  question of defining simulations in Iris is surprisingly
  interesting, and that correctness proofs of transformations of
  general recursive functions require coinductive reasoning. Clément
  Allain wrote the bulk of the correctness proof at this point, and
  finished the mechanization work over 2023.

  Finally, the present research article itself was written by Clément
  Allain and Gabriel Scherer in Summer 2023 and Spring 2024, with
  excellent review comments from François Pottier, the POPL'25
  anonymous reviewers, and Anton Lorenzen.
\end{acks}


%% file: more-ocaml.tex
\section{More on TMC in \OCaml}
\label{app:more-ocaml}

\subsection{Alternatives}
\label{subsec:alternative-impls}

We mention other approaches than TMC to solve the problem of overflowing the call stack, and explain why they were less suitable for \OCaml. We also discuss a significant implementation change between \OCaml 4 and \OCaml 5.

\subsubsection{Doing a CPS transformation}

Instead of a program transformation in \emph{destination-passing} style, we could perform a more general program transformation that can make more functions tail-recursive, for example a generic \emph{continuation-passing} style (CPS) transformation.

We have three arguments for implementing the TMC transformation:
\begin{itemize}
\item The TMC transformation generates more efficient code, using mutation instead of function calls. On the \OCaml runtime, the difference is a large constant factor.\footnote{On a toy benchmark with large-sized lists, the CPS version is 100\% slower and has 130\% more allocations than the non-tail-recursive version.}

\item The CPS transformation can be expressed at the source level, and can be made reasonably nice-looking using some monadic-binding syntactic sugar. TMC can only be done by the compiler, or using safety-breaking features.
\end{itemize}

\subsubsection{Lazy data}

Lazy (call-by-need) languages will also often avoid running into stack overflows: as soon as a lazy data structure is returned, which is the default, functions such as \ocaml{map} will return immediately, with recursive calls frozen in a lazy thunk, waiting to be evaluated on-demand as the user traverses the result structure.
Users still need to worry about tail-recursion for their strict functions; strict functions are often preferred when writing efficient code.

\subsubsection{Unlimiting the system stack}

Some operating systems can provide an unlimited system stack; such as \texttt{ulimit -s unlimited} on Linux systems -- the system stack is then resized on-demand.
Frustratingly, unlimited stacks are not available on all systems, and not the default on any system in wide use.
Convincing all users to setup their system in a non-standard way would be \emph{much} harder than performing a program transformation or accepting the CPS overhead for some programs.

\subsubsection{Using another stack}

Using the native system stack is a choice of the \OCaml~4 implementation.
Some other implementations of functional languages, such as SML/NJ, use a different stack (the \OCaml bytecode interpreter does this), or directly allocate stack frames on their GC-managed heap.
This approach can make ``stack overflow'' go away completely, and it also makes it very simple to implement stack-capture control operators, such as continuations, or other stack operations such as continuation marks.

On the other hand, using the native stack brings compatibility benefits (coherent stack traces for mixed \OCaml+C programs), and seems to improve the performance of function calls (on benchmarks that are only testing function calls and return, such as Ackermann or the naive Fibonacci, \OCaml can be 4x, 5x faster than SML/NJ.)

\paragraph{\OCaml~5}

\OCaml~5.0.0 was released in December 2022, about a year after we landed our TMC implementation in the \OCaml~4 compiler.
\OCaml~5 uses a different runtime to support multicore programming, and a different calling convention to support algebraic effects.
In particular, it only uses the system stack for C calls, and a ``cactus stack'' for \OCaml calls.

\OCaml~5 manages its own stack, but the implementors still decided to have a stack limit.
It is sensibly higher by default (1Gb at the time of writing) than the system stack (8Mb), so many ``small'' can now use non-tail-recursive functions without fears of stack overflows, but overflows remain possible and likely in practice on large inputs.
The reason to keep a (user-settable) limit for the \OCaml call stack is usability in the face of buggy programs.
When programmers write recursive functions, they occasionally go through incorrect versions with a faulty base case that fail to terminate.
In this case, we want the system to fail quickly with an error, instead of remaining silent for a few minutes, crashing once all the machine memory has been consumed.

We were curious about whether users would still see a need for TMC in \OCaml~5, or whether they would stop using the feature in practice.
The feedback we got from expert users is that stack overflows is still an issue they worry about.
We observe that TMC adoption in \OCaml code bases keep growing after the transition to \OCaml~5.

\subsection{Design Choices}
\label{subsec:design-choices}

\subsubsection{Opt-in Rather than Opt-out}\label{subsec:optin}
We made our transformation \emph{opt-in}, the user has to explicit use
the \ocaml{[@tail_mod_cons]}, rather than \emph{opt-out}, where the
compiler would automatically detect program in the TMC fragment to
transform them. There are at least three reasons for this choice, in
increasing order of importance. First, the transformation duplicates
code, as it generates a direct- and a destination-passing-style
version of the function; in particular we could suffer from
a code-size blow-up on nested transformable functions. Second, the
transformation may change the evaluation order of function arguments,
in a way that conforms to the OCaml specification but may break
programs in practice (bad!), so we would have to restrict it to only
transform calls with syntactically-pure arguments. Third and most
importantly, our implementation may have bugs, which may silently
introduce miscompilations in user code. There are millions of lines of
OCaml code out there, and we have no idea which portion would be
transformed. Doing this implicitly would risk running into
compile-time, runtime or correctness issues that we have not foreseen,
it would have been much harder to convince OCaml maintainers to
include this feature if it was opt-out.

\subsubsection{Resolving Non-determinism}

The main question faced by an implementation is how to resolve the
inherent non-determinism in the rewrite relation -- how to decide
between several possible results of the transformation. We identified
fairly different kinds of non-determinism.

\paragraph{Choices with an obviously better alternative.}
Some choices offer an alternative that is obviously better than the others, because it allows to strictly improve more programs. For example, some implementations of TMC limit themselves to calls that are immediately inside a constructor: they would optimize the recursive call to \ocaml|f| in \ocaml|Cons(x, f y)| but not, for example, in \ocaml|Cons(x1, Cons(x2, f y))|, or in \ocaml|Cons (x, if p then f y else z)|.

In the terms of \cref{subsec:specification}, those implementations only allow to optimize calling contexts of the form $\datalangTailCtx{[\datalangConsFrame]}$ or $\datalangTailCtx{[\datalangConsCtx]}$. Our implementation supports the more general situation $\datalangTMCCtx = (\datalangTailFrame | \datalangConsFrame)^{*}$. Both approaches are included in our non-deterministic relation -- we prove them both correct -- but optimizing more calls is obviously better.

\paragraph{Choices with a subtly better alternative.} In some cases it
is possible to put a bit more work in the choice heuristics, to
generate slightly better code, in terms of performance or
readability. For example, some natural way to simplify the
implementation would result in more subterms being transformed in
destination-passing-style, only to finally use the
\RefTirName{DPSBase} rule without any DPS function call. The generated
code is slightly less pleasant to read, and in our experience it can
be noticeably slower. (We detail such a situation in \cref{subsec:constructor-compression}.)

Code readability is important in our context of an on-demand program
transformation used by performance experts: those experts will often
read the intermediate representations produced by the compiler to
check that their performance assumptions hold.

\paragraph{Incomparable choices: force the user to decide} \label{subsec:user-control}
The remaining choices are between incomparable alternatives, that could
each be better than the other depending on the specific program or
programmer intent. Consider again our binary tree example,
\ocaml|Node(map f left, map f right)|: we could make either the first
or the second argument a tail-call.

Our policy in such cases is to let the user decide, by providing
control over the transformation choices using \ocaml{[@tailcall]}
program annotations, and to force the user to decide by raising an
error in case of ambiguity:
\begin{Ocaml}
  | Cifthenelse(cond, ifso, ifnot) ->
      Cifthenelse(cond, map_tail f ifso, map_tail f ifnot)
      /[^^^^^^^^^^^^^^^^^^^^^^^^^^^^^^^^^^^^^^^^^^^^^^^^^^^^]/
/[Error]/: "[@tail_mod_cons]": this constructor application may be TMC-transformed
       in several different ways. Please disambiguate by adding an explicit
       "[@tailcall]" attribute to the call that should be made tail-recursive,
       or a "[@tailcall false]" attribute on calls that should not be
       transformed.
\end{Ocaml}

This approach is incompatible with a view of the TMC
transformation as an implicit optimization, applied whenever
possible -- in that case one would rather have the compiler make
arbitrary choices rather than fail. We rather view TMC as a tool for
expert users to better reason about program performance. It should be
predictable and flexible (let the user express
their intent). Failing due to the lack of annotations is an acceptable way
to drive user interaction.

\subsubsection{First-order Implementation}\label{subsubsec:first-order}
A notable limitation of the \OCaml implementation is that it is first-order and non-modular in nature: only direct calls to known functions can be converted into DPS style, and the availability of a DPS variant is not exposed through module abstractions.
It would be possible to allow DPS calls through external modules or higher-order function parameters, by annotating interfaces and function arguments with a \ocaml{[@tail_mod_cons]} annotation, and elaborating them into a pair of functions, the direct-style and the DPS version.

\subsection{Implementation history}
\label{subsec:PR-history}

We first proposed adding TMC as an optional program transformation to
the \OCaml compiler in
May 2015.\footnote{\nonanon{URL}{\url{https://github.com/ocaml/ocaml/pull/181}}}
The proposal was received favorably, but it never received an in-depth
review and a detailed performance evaluation and remained unmerged for
years.

We restarted the implementation in 2020, resulting in a new pull July 2020 request with a modified implementation, and a careful performance
evaluation\footnote{\nonanon{URL}{\url{https://github.com/ocaml/ocaml/pull/9760}}}. The
new implementation put a larger focus on producing readable code,
giving more control to the user through annotations. It also removed
an optimization of the previous implementation that would specialize
the DPS version, generating a distinct definition for each block
offset. The new design was carefully reviewed by Basile Clément and
Pierre Chambart, and was finally merged in November 2021, available to
users with \OCaml~4.14, released in March 2022.

The TMC transformation is that it adds extra parameters to functions
(two parameters, the block and the offset). At the time the \OCaml
compiler had a limitation on some supported architectures, where it
would not optimize tail calls above a certain number of arguments
(enough that they cannot be all passed by registers), breaking the
tail-call promises of TMC on those systems. Xavier Leroy implemented
a change to the \OCaml calling convention for those architectures in
May 2021,\footnote{\url{https://github.com/ocaml/ocaml/pull/10595}},
which was motivated by the TMC work.

\subsection{Implementation}
\label{subsec:implementation}

Implementing the TMC transform is not an obvious top-down or bottom-up traversal, because it relies on information flowing in two opposite directions: the transformation is \emph{possible} for subterms that are in tail-or-constructor position relative to the root of the function (top-down information), and it is \emph{desirable} for subterms whose leaves contain calls to TMC-transformed functions (bottom-up information). A naive approach is to perform a recursive traversal that tracks the top-down context and, on each subterm, recursively traverses the whole subterm to check desirability; this has quadratic time complexity, which is best avoided in production compilers.

Instead of trying to transform each subterm in a single pass, our implementation computes for each subterm a ``choice'', a summary of all the bottom-up information that is relevant to choose how to transform this subterm -- once we have the top-down information available. This includes (1) the direct-style transform of the subterm, (2) the DPS-style transform of the subterm, (3) metadata tracking whether the transformation is beneficial (if a TMC-function call was found in tail-modulo-cons position), the list of TMC calls it contains (for error diagnostics), and whether some were explicitly requested by the user (for disambiguation). Computing transformed terms is not compositional; computing choices is compositional. We compute a choice for the whole function body (in one pass), from which we can directly extract the direct-style and DPS-style transformations of the function.

\paragraph{An applicative functor}
\label{subsec:applicative-impl}

The computation of those choices described in \cref{subsec:implementation} can be expressed elegantly. Instead of a monomorphic type \ocaml|Choice.t| that represents a choice of how to transform a term, we use a polymorphic type \ocaml|'a Choice.t| that represents how to transform zero, one or several subterms that occur in the source; for example transforming two subterms in the same context relies on a choice of type \ocaml|(lambda * lambda) Choice.t|, where \ocaml|lambda| is the type of terms in the intermediate representation we are working on. This \ocaml|'a Choice.t| type can be equipped with an applicative functor interface:

\begin{minipage}{0.4\linewidth}
\begin{Ocaml}
module Choice : sig
  type 'a t = {
    dps : 'a Dps.t;
    direct : unit -> 'a;
    tmc_calls : tmc_call_info list;
    benefits_from_dps : bool;
    explicit_tailcall_request : bool;
  }

end
\end{Ocaml}
\end{minipage}
\hfill
\begin{minipage}{0.5\linewidth}
\begin{Ocaml}
  (* construct a choice from an arbitrary term *)
  val lambda : lambda -> lambda t

  (* applicative functor interface *)
  val unit : unit t
  val map : ('a -> 'b) -> 'a t -> 'b t
  val pair : 'a t * 'b t -> ('a * 'b) t

  (* extract the direct-style and DPS transforms *)
  val direct : lambda t -> lambda
  val dps :
    lambda t -> tail:bool -> dst:offset dst ->
    lambda
\end{Ocaml}
\end{minipage}

With this interface, the easy cases of the transformation can be expressed compactly:
\begin{Ocaml}
  let rec choice ctx ~tail t =
    match t with
    [...]
    | Lifthenelse (l1, l2, l3) ->
        let l1 = traverse ctx l1 in
        let+ l2 = choice ctx ~tail l2
        and+ l3 = choice ctx ~tail l3
        in Lifthenelse (l1, l2, l3)
    [...]
\end{Ocaml}
The binding operators \ocaml|let+|, \ocaml|and+| are standard syntax for applicative-like structures in \OCaml: \ocaml|let+| desugars to \ocaml|Choice.map| and \ocaml|and+| desugars to \ocaml|Choice.pair|. \ocaml|traverse| performs a direct-style translation, and \ocaml|traverse_letrec| may also transform \ocaml|let|-bound functions marked with \ocaml|[@tail_mod_constr]| and add them to the local transformation environment.

\subsection{Evaluation: adoption}
\label{subsec:adoption}

The TMC transformation was first made available to users in \OCaml~4.14.0, released in March 2022.
At the time there were no users of the feature.
Notably, the \OCaml standard library had intentionally not been modified to start using it: before the release, we did not want the stdlib code base to depend on a not-available-yet feature, and after the release we decided not to push for adoption (we may be biased about the benefits/importance of TMC), and let other contributors propose its use, after doing their own evaluation of performance impact and code-clarity benefits.

Over time, \ocaml|[@tail_mod_cons]| was adopted in a few places in the \OCaml standard library, either to make some functions tail-recursive that previously were not, or simplify some complex tail-recursive implementations. As of spring 2024, the feature is now used in \ocaml|Hashtbl.find_all| and in \ocaml|List| module (\ocaml|init, map, mapi, map2, find_all, filteri, filter_map, take, take_while, of_seq|). \ocaml|init| and the \ocaml|map*| functions, which are the most heavily used, were unrolled once -- the minimal amount observed to preserve the non-tailrec performance for small lists. Other functions are written in the simplest possible way. (All \texttt{stdlib} uses so far build lists rather than another datatype.)

We performed a systematic search for \ocaml|[@tail_mod_cons]| usage on November 2023, and we found that it was also used
\begin{itemize}
\item in a few utility modules (general-purpose functions designed to extend or replace the standard library), notably in Jane Street's \texttt{base} library.
\item in the middle of domain-specific user code, to build lists, in 14 different projects
\item in the middle of user code, to build other types than lists, in three projects:
  \begin{itemize}
  \item in \texttt{RedPRL}\footnote{\url{https://github.com/RedPRL/ocaml-bwd/blob/fbf496b29532085b38073eaa62ba3d22ac619d5d/src/BwdNoLabels.ml\#L60}},
    it is used to build snoc lists
  \item in a project named \texttt{PL-reading-group}\footnote{\url{https://github.com/Skyb0rg007/PL-Reading-Group/blob/7002ae35ba69fb9010e5049eee88ab475bc79ec3/effects/lib/free/tseq.ml\#L44}},
    it is used to build a GADT of difference lists
  \item in \texttt{acutis}\footnote{\url{https://github.com/johnridesabike/acutis/blob/4113fb516d9c5dd6f2dbfd9657f52c5ae7a5dcae/lib/matching.ml\#L161}},
    in a group of mutually-recursive functions that builds a complex AST structure used options and nested records
  \end{itemize}
\end{itemize}

We also found three cases where the annotation is (in our opinion) misused: in two cases, the function is already tail-recursive, so there is no need for the annotation, and in a third case\footnote{\url{https://github.com/gridbugs/llama/blob/96d1c96c31ff136f465b5f37c981fff591bac6fd/src/midi/byte_array_parser.ml\#L50}} we consider that the use is gratuitous and can be replaced by already-available standard library functions.


%% file: ocaml4-appendix.tex
\section{Benchmark results on \OCaml~4}

\begin{figure}[tp]
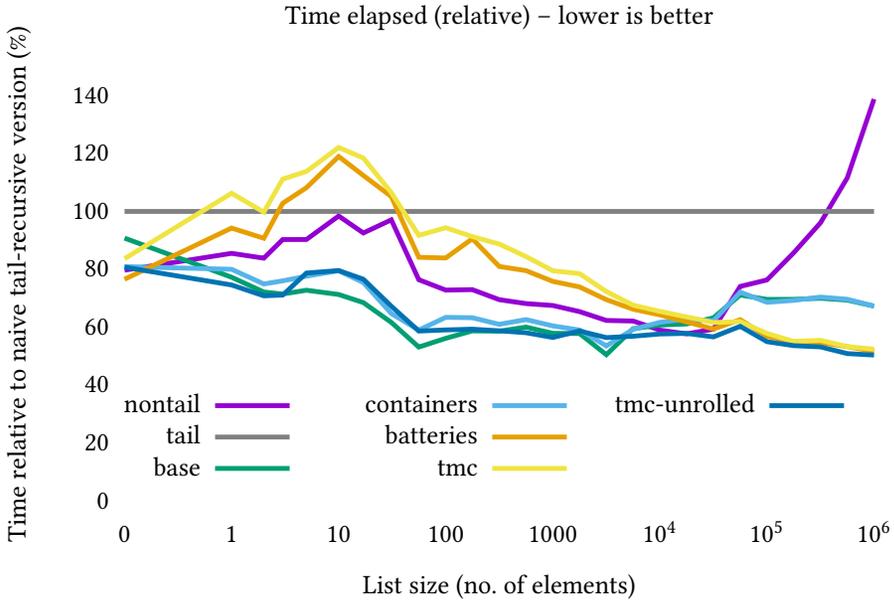

\def\svgscale{0.8}
\graphicspath{{plots/}}

\caption{\ocaml|List.map| benchmark on \OCaml~4.14}
\label{fig:bench4}
\end{figure}

The benchmark results for \OCaml~4 are given in \cref{fig:bench4}. The
results are very close to \cref{fig:bench5}, and in particular the
qualitative analysis of the results is mostly unchanged. (We used
\texttt{ulimit -s unlimited} to disable the system stack limit, to run
the non-tail-recursive version on large lists.)

%% file: plots/plot.4.pdf_tex
\begingroup%
  \makeatletter%
  \providecommand\color[2][]{%
    \errmessage{(Inkscape) Color is used for the text in Inkscape, but the package 'color.sty' is not loaded}%
    \renewcommand\color[2][]{}%
  }%
  \providecommand\transparent[1]{%
    \errmessage{(Inkscape) Transparency is used (non-zero) for the text in Inkscape, but the package 'transparent.sty' is not loaded}%
    \renewcommand\transparent[1]{}%
  }%
  \providecommand\rotatebox[2]{#2}%
  \newcommand*\fsize{\dimexpr\f@size pt\relax}%
  \newcommand*\lineheight[1]{\fontsize{\fsize}{#1\fsize}\selectfont}%
  \ifx\svgwidth\undefined%
    \setlength{\unitlength}{450bp}%
    \ifx\svgscale\undefined%
      \relax%
    \else%
      \setlength{\unitlength}{\unitlength * \real{\svgscale}}%
    \fi%
  \else%
    \setlength{\unitlength}{\svgwidth}%
  \fi%
  \global\let\svgwidth\undefined%
  \global\let\svgscale\undefined%
  \makeatother%
  \begin{picture}(1,0.66666667)%
    \lineheight{1}%
    \setlength\tabcolsep{0pt}%
    \put(0,0){\includegraphics[width=\unitlength,page=1]{plot.4.pdf}}%
    \put(0.12473333,0.10441667){\makebox(0,0)[rt]{\lineheight{1.25}\smash{\begin{tabular}[t]{r}0\end{tabular}}}}%
    \put(0,0){\includegraphics[width=\unitlength,page=2]{plot.4.pdf}}%
    \put(0.12473333,0.16436667){\makebox(0,0)[rt]{\lineheight{1.25}\smash{\begin{tabular}[t]{r}20\end{tabular}}}}%
    \put(0,0){\includegraphics[width=\unitlength,page=3]{plot.4.pdf}}%
    \put(0.12473333,0.22431667){\makebox(0,0)[rt]{\lineheight{1.25}\smash{\begin{tabular}[t]{r}40\end{tabular}}}}%
    \put(0,0){\includegraphics[width=\unitlength,page=4]{plot.4.pdf}}%
    \put(0.12473333,0.28428333){\makebox(0,0)[rt]{\lineheight{1.25}\smash{\begin{tabular}[t]{r}60\end{tabular}}}}%
    \put(0,0){\includegraphics[width=\unitlength,page=5]{plot.4.pdf}}%
    \put(0.12473333,0.34423333){\makebox(0,0)[rt]{\lineheight{1.25}\smash{\begin{tabular}[t]{r}80\end{tabular}}}}%
    \put(0,0){\includegraphics[width=\unitlength,page=6]{plot.4.pdf}}%
    \put(0.12473333,0.40418333){\makebox(0,0)[rt]{\lineheight{1.25}\smash{\begin{tabular}[t]{r}100\end{tabular}}}}%
    \put(0,0){\includegraphics[width=\unitlength,page=7]{plot.4.pdf}}%
    \put(0.12473333,0.46413333){\makebox(0,0)[rt]{\lineheight{1.25}\smash{\begin{tabular}[t]{r}120\end{tabular}}}}%
    \put(0,0){\includegraphics[width=\unitlength,page=8]{plot.4.pdf}}%
    \put(0.12473333,0.52408333){\makebox(0,0)[rt]{\lineheight{1.25}\smash{\begin{tabular}[t]{r}140\end{tabular}}}}%
    \put(0,0){\includegraphics[width=\unitlength,page=9]{plot.4.pdf}}%
    \put(0.14105,0.06941667){\makebox(0,0)[t]{\lineheight{1.25}\smash{\begin{tabular}[t]{c}0\end{tabular}}}}%
    \put(0,0){\includegraphics[width=\unitlength,page=10]{plot.4.pdf}}%
    \put(0.25326667,0.06941667){\makebox(0,0)[t]{\lineheight{1.25}\smash{\begin{tabular}[t]{c}1\end{tabular}}}}%
    \put(0,0){\includegraphics[width=\unitlength,page=11]{plot.4.pdf}}%
    \put(0.36548333,0.06941667){\makebox(0,0)[t]{\lineheight{1.25}\smash{\begin{tabular}[t]{c}10\end{tabular}}}}%
    \put(0,0){\includegraphics[width=\unitlength,page=12]{plot.4.pdf}}%
    \put(0.4777,0.06941667){\makebox(0,0)[t]{\lineheight{1.25}\smash{\begin{tabular}[t]{c}100\end{tabular}}}}%
    \put(0,0){\includegraphics[width=\unitlength,page=13]{plot.4.pdf}}%
    \put(0.58993333,0.06941667){\makebox(0,0)[t]{\lineheight{1.25}\smash{\begin{tabular}[t]{c}1000\end{tabular}}}}%
    \put(0,0){\includegraphics[width=\unitlength,page=14]{plot.4.pdf}}%
    \put(0.70215,0.06941667){\makebox(0,0)[t]{\lineheight{1.25}\smash{\begin{tabular}[t]{c}$10^4$\end{tabular}}}}%
    \put(0,0){\includegraphics[width=\unitlength,page=15]{plot.4.pdf}}%
    \put(0.81436667,0.06941667){\makebox(0,0)[t]{\lineheight{1.25}\smash{\begin{tabular}[t]{c}$10^5$\end{tabular}}}}%
    \put(0,0){\includegraphics[width=\unitlength,page=16]{plot.4.pdf}}%
    \put(0.92658333,0.06941667){\makebox(0,0)[t]{\lineheight{1.25}\smash{\begin{tabular}[t]{c}$10^6$\end{tabular}}}}%
    \put(0.0373,0.33681666){\rotatebox{90}{\makebox(0,0)[t]{\lineheight{1.25}\smash{\begin{tabular}[t]{c}Time relative to naive tail-recursive version (\%)\end{tabular}}}}}%
    \put(0.53381666,0.01691667){\makebox(0,0)[t]{\lineheight{1.25}\smash{\begin{tabular}[t]{c}List size (no. of elements)\end{tabular}}}}%
    \put(0.22076667,0.2037){\makebox(0,0)[rt]{\lineheight{1.25}\smash{\begin{tabular}[t]{r}nontail\end{tabular}}}}%
    \put(0,0){\includegraphics[width=\unitlength,page=17]{plot.4.pdf}}%
    \put(0.22076667,0.1712){\makebox(0,0)[rt]{\lineheight{1.25}\smash{\begin{tabular}[t]{r}tail\end{tabular}}}}%
    \put(0,0){\includegraphics[width=\unitlength,page=18]{plot.4.pdf}}%
    \put(0.22076667,0.1387){\makebox(0,0)[rt]{\lineheight{1.25}\smash{\begin{tabular}[t]{r}base\end{tabular}}}}%
    \put(0,0){\includegraphics[width=\unitlength,page=19]{plot.4.pdf}}%
    \put(0.51126667,0.2037){\makebox(0,0)[rt]{\lineheight{1.25}\smash{\begin{tabular}[t]{r}containers\end{tabular}}}}%
    \put(0,0){\includegraphics[width=\unitlength,page=20]{plot.4.pdf}}%
    \put(0.51126667,0.1712){\makebox(0,0)[rt]{\lineheight{1.25}\smash{\begin{tabular}[t]{r}batteries\end{tabular}}}}%
    \put(0,0){\includegraphics[width=\unitlength,page=21]{plot.4.pdf}}%
    \put(0.51126667,0.1387){\makebox(0,0)[rt]{\lineheight{1.25}\smash{\begin{tabular}[t]{r}tmc\end{tabular}}}}%
    \put(0,0){\includegraphics[width=\unitlength,page=22]{plot.4.pdf}}%
    \put(0.80176667,0.2037){\makebox(0,0)[rt]{\lineheight{1.25}\smash{\begin{tabular}[t]{r}tmc-unrolled\end{tabular}}}}%
    \put(0,0){\includegraphics[width=\unitlength,page=23]{plot.4.pdf}}%
    \put(0.53381667,0.60656667){\makebox(0,0)[t]{\lineheight{1.25}\smash{\begin{tabular}[t]{c}Time elapsed (relative) – lower is better\end{tabular}}}}%
  \end{picture}%
\endgroup%